\documentclass[nofootnoteinbib]{article}
\pdfoutput=1
\usepackage{jcappub}

\usepackage[english]{babel}
\usepackage[margin=2.54cm]{geometry}
\usepackage{lmodern}

\usepackage[T1]{fontenc}
\usepackage[utf8]{inputenc}
\usepackage{color}
\usepackage{amsmath}
\usepackage{amssymb}
\usepackage{esint}
\usepackage{xcolor}
\usepackage{graphicx}

\usepackage{calligra}

\usepackage{breakurl}

\definecolor{madecolor}{rgb}{1.,0.6,0.2}
\hypersetup{
colorlinks=true,
citecolor=madecolor,
linkcolor=madecolor,
urlcolor=madecolor
}

\newcommand{\be}{\begin{equation}} 
\newcommand{\ee}{\end{equation}}
\newcommand{\bea}{\begin{eqnarray}}
\newcommand{\eea}{\end{eqnarray}}
\newcommand{\nn}{\nonumber}

\newcommand\osq{\overline{\square}}
\newcommand\ovr{\overline{R}}

\newcommand\ovX{\overline{X}}
\newcommand\ovf{\overline{f}}
\newcommand\ovU{\overline{U}}

\newcommand{\dd}{{\rm d}}
\newcommand{\lcdm}{\Lambda {\rm CDM}}
\newcommand{\hc}{\mathcal{H}}
\newcommand{\hcl}{h_\Lambda}
\newcommand{\dwlcdm}{{\rm DW}_{\Lambda \textsf{CDM fid}}}

\usepackage{blindtext}
\usepackage{bbding}

\date{\today}

\title{Observational Constraints in Nonlocal Gravity: the Deser-Woodard Case\footnote{Based on observations obtained with Planck (http://www.esa.int/Planck), an ESA science mission with instruments and contributions directly funded by ESA Member States, NASA, and Canada.}}

\author[a]{Luca Amendola}
\emailAdd{l.amendola@thphys.uni-heidelberg.de}
\author[b,c]{Yves Dirian\textsuperscript{\FourClowerOpen}}
\emailAdd{yves.dirian@ics.uzh.ch}
\author[a]{Henrik Nersisyan}
\emailAdd{h.nersisyan@thphys.uni-heidelberg.de}
\author[d]{Sohyun Park}
\emailAdd{park@fzu.cz}

\affiliation[a]{Institut f\"ur Theoretische Physik, Ruprecht-Karls-Universit\"at Heidelberg, Philosophenweg 16, 69120 Heidelberg, Germany}
\affiliation[b]{Department of Theoretical Physics and Center for Astroparticle Physics, University of Geneva, Quai Ansermet 24, CH--1211 Gen\`eve 4, Switzerland}
\affiliation[c]{Center for Theoretical Astrophysics and Cosmology, Institute for Computational Science, University of Z\"urich, Winterthurerstrasse 190, CH--8057 Z\"urich, Switzerland}
\affiliation[d]{CEICO, Institute of Physics of the Czech Academy of Sciences, Na Slovance 2, 182 21 Praha 8, Czechia}

\abstract{
We study the cosmology of a specific class of nonlocal model of modified gravity, the so--called Deser--Woodard (DW) model, modifying the Einstein--Hilbert action by a term $\sim R f(\Box^{-1}R)$, where $f$ is a free function. Choosing $f$ so as to reproduce the $\lcdm$ cosmological background expansion history within the nonlocal model, we implement the model in a cosmological linear Einstein--Boltzmann solver and study the deviations to GR the model induces in the scalar and tensor perturbations. We observe that the DW nonlocal model describes a modified propagation for the gravitational waves, as well as a lower linear growth rate and a stronger lensing power as compared to $\lcdm$, up to several percents. Such prominent growth and lensing features lead to the inference of a significantly smaller value of $\sigma_8$ with respect to the one in $\lcdm$, given \textit{Planck} CMB+lensing data. The prediction for the linear growth rate $f \sigma_8$ within the DW model is therefore significantly smaller than the one in $\lcdm$ and the addition of growth rate data $f \sigma_8$ from Redshift-space distortion measurements to \textit{Planck} CMB+lensing, opens a (dominant) tension between Redshift-space distortion data and the reconstructed \textit{Planck} CMB lensing potential. However, model selection issues only result in ``weak'' evidences for $\lcdm$ against the DW model given the data. Such a fact shows that the joined datasets we consider are not constraining enough for distinguishing between the models on firm grounds. As we discuss, the addition of galaxy WL data or the consideration of cosmological constraints from future galaxy clustering, weak lensing surveys, but also third generation gravitational wave interferometers, prove to be useful for discriminating modified gravity models such as the DW one from $\lcdm$, within the close future.}

\begin{document}
\maketitle
\flushbottom

%%%%%%%%%%%%%%%%%%%%%%%%%%%%%%%%%%%%%
\section{Introduction}\label{sec:intro}
%%%%%%%%%%%%%%%%%%%%%%%%%%%%%%%%%%%%%

The observations of a variety of complementary cosmological probes such as distant Type Ia supernovae (SNIa), the Cosmic Microwave Background (CMB) or the clustering properties [e.g. Baryonic Acoustic Oscillations (BAO), Redshift-Space Distortions (RSD)] and the weak lensing (WL) of galaxies have provided data of unpreceding quality. Given these data, the inferred constraints on the parameter space of the current standard model of cosmology $\lcdm$, have reached exquisite accuracy -- up to percent-level in the case of its six-dimensional ``minimal'' or ``base'' cosmological parametrisation (see e.g. Refs.~\cite{Alam:2016hwk,Abbott:2017wau,Aghanim:2018eyx}). As a consequence, the base $\lcdm$ model has been shown to be able to explain these data with high significance and consistently within each probe (internal) and also when the latter are joined together (external, also known as concordance).

Despite such astonishing capabilities, the model still suffers from theoretical flaws as well as observational weaknesses that obscure its credibility on fundamental theoretical and empirical grounds. 
On the theoretical side, $\Lambda$ is a simple, dimensionful number that lacks justifications about its nature and late-time domination (see for example Ref.~\cite{Weinberg1989,Padmanabhan:2002ji,Carroll:2003qq,Bianchi:2010uw} for reviews).
On the observational one, A. Riess \textit{et al.} recently reported in Ref.~\cite{Riess2018} that the direct measurement of $H_0$ from nearby SNIa is significantly larger (at $3.5\sigma$) than the value inferred from the  \textit{Planck} 2018 CMB analysis of Ref.~\cite{Aghanim:2018eyx}, which is also more model--dependent. Other local determinations of $H_0$ such as the ones of Refs.~\cite{EfstaH02013,2016arXiv160701790B,Riess:2016jrr,Cardona:2016ems} also generically prefer values higher than the one pinpointed by \textit{Planck}, but see Ref.~\cite{Gomez-Valent:2018hwc} for a different conclusion. In addition, cosmic shear measurements in the $\sigma_8$--$\Omega_M$ plane from weakly lensed galaxy maps collected by the \textit{CFHTLenS} \cite{2012MNRAS.427..146H} or \textit{KiDS-450} \cite{Kuijken:2015vca} survey were also shown to display a tension with the constraints inferred by \textit{Planck} \cite{Planck2015CP}. The same remark also potentially applies to $\sigma_8$--$\Omega_M$ constraints given non-\textit{Planck} (e.g. X--ray) and \textit{Planck}--\textit{SZ} selected cluster counts \cite{Ade:2015fva,Aghanim:2018eyx}. However, all these empirical results are still possibly driven by uncontrolled systematics and refined analyses of the data are first needed before meaningful conclusions can be established. Indeed, this fact is well illustrated by the recent controvercy triggered by the low-valued distance scale estimate of $H_0$ from Ref.~\cite{Shanks:2018rka}, which proposed an alternative pipeline including the use of \textit{GAIA DR2} ``quasar''-corrected parallaxes as compared with the work of Ref.~\cite{Riess2018} (see also Ref.~\cite{Riess:2018kzi,Shanks:2018dsp} for further discussions)\footnote{The significance of the high--$z$ (CMB) and low--$z$ tension on $H_0$--derived inferences however mildens when building an inverse distance ladder \cite{Cuesta:2014asa}. This fixes the comoving sound horizon $r_s$ at the drag epoch using joined low--z BAO and SNIa distance scale data which can then be compared with independent inferences from CMB observations \cite{Aylor:2018drw}.}. Another example comes from Ref.~\cite{Troxel:2018qll}, which improved the modelisation of the noise covariance within the data analysis pipeline of the \textit{KiDS-450} survey and increased the agreement between galaxy WL and \textit{Planck} CMB. Moreover, comparing cluster counts constraints between \textit{Planck} and external experiments requires the knowledge of the relation between the X--ray mass and the total cluster mass, which is currently parametrized by an approximate hydrostatic mass linear bias (see for example Refs.~\cite{Ade:2015fva,Bolliet:2017lha,Aghanim:2018eyx}).  

These facts support the development of phenomenological models of dark energy/modified gravity (see for instance Refs.~\cite{ruiz2010,Clifton:2011jh,Joyce:2014kja,Berti:2015itd,Ishak:2018his} for reviews and Ref.~\cite{Amendola2010} for a standard textbook) that aim to explain better current data and, at the same time, can provide hints for solutions to the problems of $\Lambda$. Modified gravity models all have a common thread in that they evade Lovelock's theorem \cite{Lovelock:1971yv,Lovelock:1972vz} and, given that General Relativity (GR) is well-tested on solar system scales, they must modify its dynamics in its infrared regime (IR). This can be realized in various fashions and examples of IR modifications are provided by scalar-tensor theories including one extra degree-of-freedom such as quintessence (see e.g. Ref.~\cite{Tsujikawa:2013fta} for a review), the so-called beyond Horndeski ``degenerate higher order scalar-tensor theories'' \cite{Langlois:2015cwa} or the more general effective field theory of dark energy \cite{Gleyzes:2013ooa,Gleyzes:2014rba}. In order to reproduce the predictions of GR on solar system scales (high density regions), scalar-tensor theories must either be effectively decoupled to baryons (see e.g. Ref.~\cite{Amendola:1999er}), or
exhibit a screening mechanism of, for example, the chameleon \cite{Khoury:2003aq,Brax:2004qh}, symmetron \cite{Hinterbichler:2010es} or Vainstein \cite{Vainshtein1972} type, suppressing their fifth force on solar system scales. When reasonably close from $\lcdm$ for the same cosmological parameter values, alternative cosmological models can then legitimately and efficiently be constrained by using the aforementioned cosmological probes. This is well illustrated by the recent works of Ref.~\cite{Durrive:2018quo} on quintessence theories and of Ref.~\cite{Kreisch:2017uet} on Horndeski theories. Furthermore, these models can also be used for forecasting cosmological constraints and model selection issues from future cosmological surveys such as \textit{Euclid} \cite{Euclid1,Euclid2}, \textit{DESI} \cite{DESI1,DESI2}, \textit{LSST} \cite{2009arXiv0912.0201L}, \textit{SKA} \cite{SKA1,SKA2} or \textit{Stage-4 CMB} experiments \cite{Abazajian:2016yjj} (see e.g. Refs. \cite{Alonso:2016suf,Casas:2017eob}). Moreover, prospects of future observations of the Gravitational Waves (GWs) produced by binary mergers from third (or $2.5$) generation interferometers such as \textit{LISA} \cite{Audley:2017drz}, the \textit{Einstein Telescope} \cite{Sathyaprakash:2009xt} or \textit{Cosmic Explorer} \cite{Evans:2016mbw}, will also provide exceptional complementary information to constrain the expansion history as well as deviations to GR, such as in the propagation properties of GWs -- in particular in the amount of their damping under the Hubble flow (see e.g. Refs.~\cite{Cai:2016sby,Caprini:2016qxs,Amendola:2017ovw,Belgacem:2018lbp,Du:2018tia} for constraints on dark energy/modified gravity using standard sirens).

Another class of IR modifications to GR consists in scalar nonlocal modifications\footnote{Scalar nonlocal modifications refer to the use of composite operators made of diffeomorphism--scalar quantities including operators that are non-polynomial in their derivatives to modify GR. As will be discussed below, on phenomenological grounds, most of the scalar nonlocal gravity theories considered to date involve combinaisons of the Ricci scalar $R$ and a Green's function of the d'Alembert operator $\Box^{-1}$, such as $\sim \Box^{-1}R, \, \sim R\Box^{-1}R, \sim \Box^{-2}R ,\, {\rm etc}$. This is opposed to tensor nonlocal modifications where higher rank tensors/operators are also included. Several models of tensor nonlocalities were shown to generically exhibit growing modes at the cosmological background or linear perturbation level, preventing them from modelling a phenomenologically viable cosmological dynamics (see e.g. Refs.~\cite{FM2013,Cusin:2015rex,Nersisyan:2016jta}). Nonetheless, exceptions might still exist (see e.g. Ref.~\cite{Tsamis:2014hra}).}. In the beginning of this century, it has been reported that such corrections can be induced from higher dimensions such as within the Dvali-Gabadadze-Porrati (DGP) braneworld model \cite{2000PhLB..485..208D,Dvali:2000xg}, but the model has been shown not to be phenomenologically viable because of the presence of ghosts in the self-accelerating branch (see e.g. Refs.~\cite{Izumi2006,Koyama2007} and references therein). Nevertheless, it is still believed that such corrections can also result from quantum effective non-perturbative corrections. In this context, prototypical mechanisms are provided by renormalisation group flow corrections to the bare gravitational couplings, characterized by a fixed point in the ultraviolet (UV) \cite{Hamber:1994jh,Hamber:2005dw} (see Ref.~\cite{Amendola:2017qge} for a recent cosmological study), or by dynamical mass-scale generation in the IR, such as a mass for the conformal mode of the graviton as suggested in Ref.~\cite{Maggiore:2015rma} (see also Ref.~\cite{Belgacem2017a} for further details along these lines).
Quite intriguingly, the background independent lattice quantum gravity computations of Ref.~\cite{Knorr:2018kog} have recently shown a ``first-hand evidence for the presence of nonlocal terms [in the gravitational quantum effective action] which could affect the gravitational dynamics at cosmic scales''. 

Phenomenologically, the nonlocal models inspired by quantum averaging processes, such as the ones studied  in Refs.~\cite{Maggiore:2014sia,Amendola:2017qge,Belgacem2017a}, exhibit self-accelerating solutions that are generically driven by an effective dark energy component whose equation of state $w$ lies on the phantom side, i.e. $w < -1$ (see also the more exotic but related example of Ref.~\cite{Vardanyan:2017kal}). For conserved dark energy density and fixed cosmological parameter values, this implies that the Hubble expansion history at late-time described by the alternative cosmological model is reduced compared to the one predicted by $\lcdm$. As a consequence, when constrained with distance indicators such as the CMB, the alternative model will generically prefer a higher value of $H_0$ and will therefore be in better agreement with direct measurements as compared to $\lcdm$. However, if the dark energy is too phantom, a tension between CMB and distant SNIa measurements can arise.
Moreover, a smaller Hubble expansion rate at late-time also reduces the Hubble friction to matter fluctuations and therefore gives raise to a higher growth of linear as well as nonlinear structures as compared to $\lcdm$, degrading the agreement with growth data. One can however exploit the degeneracy between dark energy/modified gravity and the (absolute) neutrino mass to reestablish the compatibility of the model with the data \cite{Dirian2017} (see also Refs.~\cite{2014MNRAS.440...75B,2014PhRvD..90b3528B,Bellomo:2016xhl}). These facts are well illustrated in Refs.~\cite{Dirian:2014ara, 2014JCAP...09..031B,Dirian2017}, where the phenomenology of the so-called RR nonlocal gravity model has been analysed. This model has recently been put under extensive observational constraints \cite{Dirian:2014ara,Dirian:2014bma,Dirian:2016puz,Dirian2017,PhysRevD.98.124040} and was shown to explain CMB+BAO+SNIa+RSD data as well as $\lcdm$, when the absolute neutrino mass is left as a free parameter \cite{Dirian2017}. Furthermore, the RR model has also been used for developing future experiments' data analysis pipelines in forecasting cosmological constraints from galaxy clustering and WL surveys in Ref.~\cite{Casas2019:inprep} and from third generation GWs interferometers  in Ref.~\cite{Belgacem:2018lbp}.   

A nonlocal model that has %became
become popular in the past decade has been proposed by  S.~Deser and R.~Woodard in Ref.~\cite{Deser2007}. In the Deser--Woodard (DW) model, GR is modified by an extra term of the form $R f(\Box^{-1} R)$ to the Einstein--Hilbert action, where $f$ is a dimensionless free function. As for $f(R)$ theories, this model has no predictive power as long as the function $f$ is left unspecified. However, it has been shown in Ref.~\cite{Deffayet2009} (whose results are reproduced in Sec.~\ref{sec:model} below), that once a given $\lcdm$ Hubble expansion history is specified as $H_{\Lambda {\rm CDM}}(z)$, one can solve for $X\equiv \Box^{-1}R$ and $f(X)$ in terms of $H_{\Lambda {\rm CDM}}(z)$, so as to reconstruct the same $\lcdm$ expansion history within the DW model, i.e. for obtaining $H_{{\rm DW}}(z) \equiv H_{\Lambda {\rm CDM}}(z)$, at any redshift. The same reconstruction technique can also be carried out for non-standard $\lcdm$ cosmologies such as $w$CDM models \cite{Park:2016jym} and other simple choices for $f$ were explored as well in Ref.~\cite{Koivisto2008}. Once $f(X)$ is fixed in such a way, no extra freedom is left and the distinction between $\lcdm$ and DW cosmologies exclusively lies in the linear and nonlinear observables they describe. The impact on the linear growth of structures has first been studied  in Ref.~\cite{Park:2012cp, Dodelson2014}. 
As recognized  in Ref.~\cite{Park2017}, the authors of Ref.~\cite{Dodelson2014} erroneously concluded that RSD data favour $\lcdm$ over the DW model at a significance level of $\sim 8 \sigma$, because of an excess of growth described in DW. However, the work of Ref.~\cite{Nersisyan2017} conducted an equivalent analysis but at the so-called ``localized level'' in the equations of motion and concluded that, for the same cosmological parameter values, the DW model actually predicts a linear growth of structures that is weaker than in $\lcdm$, only up to several percent in the quasi-static approximation. This fact was confirmed in Ref.~\cite{Park2017}, which moreover established the equivalence between the nonlocal and localised versions on a particular set of initial conditions and the validity of the quasi-static approximation studied in Ref.~\cite{Nersisyan2017}.

In the present work, we complement the past analyses of Refs.~\cite{Nersisyan2017,Park2017} in solving the full equations of motion at background and linear perturbation levels in the scalar and tensor sectors, within the cosmological context. For doing so, we implement the DW nonlocal gravity model in a modified version of the linear Einstein--Boltzmann code CLASS \cite{CLASS} and study its cosmological phenomenology from a modified gravity perspective by analysing relevant indicators of deviations from GR. We then perform cosmological parameter inference and model selection with the Monte Carlo Markov Chain (MCMC) sampler MONTEPYTHON \cite{MP} and confront $\lcdm$ against DW given high precision CMB+SNIa+RSD data. 
The paper is organized as follows. In Sec.~\ref{sec:model}, we introduce the DW model and present the full set of modified Einstein equations needed to evolve its linear cosmological perturbations. In Sec.~\ref{sec:pheno}, we display the deviations of the DW model to GR given the same parameter values through the use of relevant indicators. Thereafter, we present the CMB+SNIa+RSD datasets we use in Sec.~\ref{sec:data} and perform observational constraints on the $\lcdm$ and DW models to which we display the inferred cosmological parameter distributions in Sec.~\ref{sec:MCMC}. We then compare both models against each other in a (approximate) Bayesian perspective. Our conclusions and future perspectives are discussed in Sec.~\ref{sec:conc}.

%%%%%%%%%%%%%%%%%%%%%%%%%%%%%%%%%%%%%
\section{Cosmology of the DW Model}\label{sec:model}
%%%%%%%%%%%%%%%%%%%%%%%%%%%%%%%%%%%%%

We present the most relevant properties of the DW model of Ref.~\cite{Deser2007} and present its FLRW cosmological background and linear perturbation evolution equations. We also reproduce the resulting prescription of Ref.~\cite{Deffayet2009}, for fixing the distortion function $f$ so as to reconstruct the $\lcdm$ FLRW background expansion history within the DW model.

\subsection{Model and Cosmological Background}

\noindent The DW model is given by the action~\cite{Deser2007},
\begin{align}\label{eq:DWaction}
S_{\mathrm{DW}} = \frac{1}{16 \pi G} \int \dd^4 x \sqrt{-g} ~ R \bigg[ 1 + f\bigg( \frac{1}{\Box}R \bigg) \bigg] \, , 
\end{align}
where $\Box^{-1}$ is the Green's function of the d'Alembert operator $\Box \equiv g^{\mu \nu} \nabla_\mu \nabla_\nu$ and $f$ is an arbitrary dimensionless function.
The modified Einstein equations are obtained by varying the action with respect to the (inverse) metric $g^{\mu \nu}$,
\begin{equation}\label{eq:einsteineq1}
G_{\mu\nu} + \Delta G_{\mu\nu}= 8 \pi G \, T_{\mu\nu} \, , 
\end{equation}
where $\Delta G_{\mu\nu}$ is the correction to Einstein's equations induced by the nonlocal distortion function $f$. The energy-momentum tensor of matter is,
\begin{align}
T_{\mu \nu}(x) \equiv -\frac{2}{\sqrt{-g}} \frac{\delta S_M}{\delta g^{\mu \nu}(x)} \,.
\end{align}
For a perfect fluid, when expressed in the frame of an observer $u^\mu$ comoving with it, the stress tensor takes the form,
\begin{align}
T_{\mu \nu} = (\rho + p) u^\mu u^\nu + p \, g_{\mu \nu} + \pi_{\mu \nu} \, , \qquad u^\mu \equiv \frac{\dd x^\mu}{\dd s} \, ,
\end{align}
where the infinitesimal $\dd s \equiv \sqrt{-\dd s^2}$ is the observer proper time, $\rho$ and $p$ are the energy and pressure density of the fluid probed by the observer, respectively, and $\pi_{\mu \nu}$ is the traceless--transverse anisotropic stress tensor whose FLRW background value vanishes.  

On a flat Friedmann-Lema\^{i}tre-Robertson-Walker (FLRW) background in conformal time $\tau$, one can write the metric as,
\begin{equation}
\dd s^{2}= a^2(\tau) \big(-\dd \tau^{2}+ \dd \vec{x}^{2} \big) \, , \qquad H \equiv \frac{\partial_\tau a}{a^2} \, , 
\end{equation}
In that setting, the components $\Delta G_{\mu\nu}$ take the form,
\begin{align}
\Delta \overline{G}_{00} &=
\frac{1}{a^2}\Bigl[ 3 a^2 H^2 + 3 a H \partial_\tau \Big]
\biggl\{f \big( \ovX \big) + \frac{1}{\osq}\Bigl[\ovr f_{,}\big( \ovX \big)\Bigr] \biggr\}
+\frac{1}{2a^2}
\partial_{\tau} \ovX \, \, \partial_{\tau}\bigg(\frac{1}{\osq}\Bigl[\overline{R} f_{,}\big( \ovX \big) \Bigr]\bigg)\;, 
\label{eq:DG00}\\
\Delta  \overline{G}_{ij} &=
\delta_{ij}\Bigg[ 
\frac{1}{2}
\partial_{\tau} \ovX \, \, \partial_{\tau}\bigg(\frac{1}{\overline{\square}}\Bigl[\overline{R} f_{,}\big( \ovX  \big) \Bigr]\bigg) \, \nonumber \\
&\hspace{4cm}- \Big( 2 a H'+ 3 a^2 H^2 + \partial_\tau^2+a H \partial_\tau\Big)
\biggl\{ f \big( \ovX \big) + \frac{1}{\osq}\Bigl[\ovr f_{,}\big( \ovX \big) \Bigr]\biggr\}
\Bigg] \, ,
\label{eq:DGij}
\end{align}
where primes $'$ denote derivatives with respect to conformal time, overbars $\, \bar{~} \,$ denote background quantities, $f,$ is the derivative of $f$ with respect to, 
\begin{align}
\bar{X} \equiv \osq^{-1} \bar{R} \, . \label{eq:auxX}
\end{align} 
In order to numerically evolve the system, it is convenient to write the equations of motion in the local form. This can be done by inverting the $\Box^{-1}$ operator in Eq.~\eqref{eq:auxX} and introducing another auxiliary field $U$ such as at the fully covariant level, 
\begin{eqnarray}
\Box X & \equiv & R \, , \label{eq:boxx} \\
\Box U & \equiv & R \, f_{,}  \, .\label{eq:boxu}
\end{eqnarray}
Given some initial spacelike hypersurface at $\tau_0$, one can for instance solve for $X$ as,
\begin{align}
X = \Box^{-1} R \equiv \int^\tau_{\tau_0} \dd^4 y \, \, G_{\rm ret}(x,y) R(y) + X_{\rm hom}(\tau , \vec{x}) \, , 
\end{align}
where the Green's function  is of the retarded kind $ G(x,y) \equiv G_{\rm ret}(x,y)$, and $X_{\rm hom}$ is the homogeneous solution $\Box X_{\rm hom} = 0$ (see e.g. Ref.~\cite{2014JCAP...10..065D, Park:2019btx} for more details).
Once $\tau_0$ and the type of the Green's function are fixed, the initial conditions of $X$ are determined by the choice of the homogeneous solution $X_{\rm hom}$ and its first derivative. As the Ricci scalar is negligible compared to the typical energy scale in the radiation dominated era (RD), i.e. $\left. | \bar{R}/H^2 | \right|_{\rm RD} \ll 1$, we consider the case where these initial conditions are vanishing\footnote{Ref.~\cite{Park:2019btx} studied the localised version on various initial conditions and found that instabilities were emerging for non--trivial initial conditions.}. 
Equations~\eqref{eq:DG00},\eqref{eq:DGij} therefore take the form,
\begin{eqnarray}
\Delta \overline{G}_{00} & = & \frac{1}{a^2}\left[(3 a^2 H^2+3 a H \partial_{\tau})(\ovf+\ovU)+\frac{1}{2}\ovX'\ovU'\right] \, , \\
\nonumber \\
\Delta \overline{G}_{ij} & = & \delta_{ij}\Bigg[\frac{1}{2}\ovX'\ovU'-(2 a H' + 3 a^2 H^2 + \partial_\tau^2 + a H \partial_\tau)(\ovf+\ovU)\Bigg] \, .
\end{eqnarray}
The cosmological dynamics of the background auxiliary fields is provided by Eqs.~\eqref{eq:boxx},\eqref{eq:boxu} on a flat FLRW background,
\begin{eqnarray}
\ovX''+2 a H \ovX' &=& -6\left(a H'+ 2 a^2 H^{2}\right) \, ,\\
\ovU''+2 a H \ovU' &=& -6\ovf_{,}\left(a H'+ 2 a^2 H^{2}\right) \, , \label{eqU}
\end{eqnarray}
and the modified Friedmann equations read, 
\begin{align}
&H^{2}\left(1+\ovf+\ovU\right)+ \frac{H}{a}(\ovf'+\ovU')+\frac{1}{6 a^2}\ovX'\ovU' = \frac{8\pi G}{3} \bar{\rho} \, , \\
&\left(1+\ovf+\ovU\right) H' = -\left\lbrace 4\pi G \, \bar{p} + \frac{3}{2} H^{2}\left(1+\ovf+\ovU\right) + \frac{H}{2a}\left(\ovf'+\ovU'\right)+\frac{1}{2a^2}\left(\ovf''+\ovU''\right)-\frac{1}{4 a^2}\ovX'\ovU'\right\rbrace \, .
\end{align}
Once the function $f(\ovX)$ is provided, these equations can be numerically integrated. As discussed above, we focus here on a particular class of DW models where $f(\ovX)$ is chosen so as the $\lcdm$ expansion history is reproduced within the DW model, for non-trivial $f$ \cite{Deffayet2009}. Before presenting the reconstruction method of Ref.~\cite{Deffayet2009}, we display the set of modified Einstein equations of the DW model within linear cosmological perturbation theory.

%%%%%%%%%%%%%%%%%%%%%%%%%%%%%%%%%%%%
\subsection{Linear Cosmological Perturbations} \label{sec:eqlin}
%%%%%%%%%%%%%%%%%%%%%%%%%%%%%%%%%%%%

We study the linear cosmological perturbations of the DW model in the scalar and tensor sectors. The perturbed FLRW metric is taken in the conformal Newtonian (longitudinal) gauge, 
\begin{equation}
\dd s^{2} = a^{2}\left[-\left(1+2\Psi\right) \dd \tau^{2}+\big[ \left(1-2\Phi\right)\delta_{ij} + h_{ij} \big] \dd x^{i}\dd x^{j}\right] \, .\label{eq:interval}
\end{equation}
where $h_{ij}$ is traceless and transverse with respect to $\partial_i$ and the adopted convention agrees with the ones of  Ma \&  Bertschinger in Ref.~\cite{Ma1995} and CLASS \cite{CLASS}. \\

\subsubsection{Scalar Sector}

In the scalar sector, the longitudinal trace of the perturbed $_{ij}$ component of Einstein equations Eq.~\eqref{eq:einsteineq1}, yields,
\begin{equation}\label{eq:psieq}
\Psi=\Phi-\frac{1}{1+\ovf+\ovU}\left[\frac{12 \pi G \, a^{2}}{k^{2}} \left(\bar{\rho}+\bar{p} \right)\sigma +\delta f  + \delta U \right] \, ,
\end{equation}
where we have defined, 
\begin{align}
\left(\bar{\rho}+\bar{p} \right)\sigma \equiv \frac{1}{a^2} \delta^{im} \delta^{jn} \bigg( \frac{\partial_i \partial_j}{\partial^2} - \frac{1}{3} \delta_{ij} \bigg) \pi_{mn} \, , 
\end{align}
with the flat Laplace operator $\partial^2 \equiv \delta^{ij} \partial_i \partial_j$, and we have written, $\delta f \equiv \ovf_{,} \delta X$. According to the integration scheme adopted  in CLASS, we then solve the longitudinal part of the perturbed $_{0i}$ component of Eq.~\eqref{eq:einsteineq1}, to obtain $\Phi '$,
\begin{align}
\Phi '=& -\mathcal{H}\Psi+\frac{1}{1+\ovf + \ovU}\Biggl[\frac{4\pi G \, a^{2}\left(\bar{p}+\bar{\rho}\right)\theta}{k^{2}} - \frac{1}{2}\left(\ovf' +\ovU' \right)\Psi \nonumber \\
&\hspace{5.5cm}+ \frac{1}{2}\left(\delta f'+\delta U'\right)-\frac{\mathcal{H}}{2}\left(\delta f+\delta U\right)-\frac{1}{4}\left(\delta X \ovU'+\delta U \ovX' \right)\Biggr],
\end{align}
where $\mathcal{H} \equiv a'/a$ and $\theta \equiv \bar{\nabla}^{i}v_{i}$. The dynamics of the linear perturbations of the auxiliary fields $\delta X$ and $\delta U$ is provided by,
\bea
\delta X'' + 2 \mathcal{H}\delta X' + k^{2} \delta X
&=& 6\Phi'' + (6\mathcal{H}+\ovX')(\Psi'+3 \Phi') -2k^{2}\left(\Psi-2\Phi \right) \, , 
 \\
\delta U'' + 2 \mathcal{H}\delta U' + k^{2} \delta U
&=& \ovf_{,} \left[6\Phi''+(6 \mathcal{H}+\ovU')(\Psi'+3 \Phi') -2k^{2}(\Psi-2\Phi)\right] - 6\ovf_{,,}\left(\mathcal{H}'+\mathcal{H}^{2}\right)\delta X \, ,
\eea
where we have used the background equations of motion to replace $\ovX''$ and $\ovU''$. To complete the above set of equations, we still need to find expressions for $\Phi ''$ and $\Psi '$. The former can be obtained from the trace of the perturbed $_{ij}$ component of Eq.~\eqref{eq:einsteineq1},
\begin{align}
\Phi'' = &\frac{1}{(1+\ovf+\ovU - 6 \ovf_{,})} \Biggl\{  4 \pi G \, a^2 \delta p + \frac{1}{6}\left(\mathcal{H}^2+6\mathcal{H}'+2k^{2}\right)\left(\delta f+\delta U\right) + \frac{1}{2}\mathcal{H}\left(\delta f'+ \delta U'\right)  
+\frac{1}{2}\left(\delta f '' -  \ovf_{,} \delta X'' \right) 
\nn \\ 
&+ \frac{\ovf_{,}}{2} \bigg[ -2 \mathcal{H} (\delta X' + \delta U'/ \ovf_{,}) - k^2 (\delta X + \delta U/ \ovf_{,}) - 4k^{2}(\Psi-2\Phi)  \nn \\
&+ (12 \mathcal{H} + \ovX' + \ovU'/ \ovf_{,}) ( \Psi'+3 \Phi')
-6\ovf_{,,} \left(\mathcal{H}'+\mathcal{H}^{2}\right)\delta X /\ovf_{,}  \bigg] - \frac{1}{2}\left(\ovX'\delta U'+\ovU' \delta X'\right)
\nn \\
&
-\frac{1}{6}\bigg[6(\mathcal{H}^2+2\mathcal{H}')(1+\ovf+\ovU)+ \frac{1}{2}\mathcal{H}(\ovf' + \ovU')+2\big(\ovf'' - 2 \mathcal{H} \ovU' - 6 \ovf_{,}(\mathcal{H}' + \mathcal{H}^2) \big)-\frac{1}{4}\ovX'\ovU' \bigg] \Psi
\nn \\
&+\frac{k^2}{3}(1+\ovf+\ovU)(\Psi-\Phi)
-\frac{1}{2}\left[2\mathcal{H}(1+\ovf+\ovU) + \ovf'+\ovU'\right](\Psi'+2\Phi') 
   \Biggr\} \, , 
\end{align}
where we have replaced the $''$ quantities by using their equations of motion and we have written,
\begin{align}
\delta f' &\equiv \ovf_{,,} \ovX' \delta X + \ovf_{,} \delta X' \, , \\
\delta f'' -  \ovf_{,} \delta X'' &\equiv \frac{\delta^2 \ovf_{,}}{\delta \tau^2} \delta X + 2 \frac{\delta \ovf_{,}}{\delta \tau} \delta X' \, , \\
\ovf_{,,} &\equiv \big( \ovX' \big)^{-1} \big[ \ovf'' \big( \ovX' \big)^{-1} - \ovf' \big( \ovX' \big)^{-2} \ovX''  \big] \, , \\
&= \ovf'' \big( \ovX' \big)^{-2} + \ovf' \big[ 2 a \hc \big( \ovX' \big)^{-2} + 6 \big( \ovX' \big)^{-3} \big( a \hc' + 2 a^2 \hc^2 \big) \big] \, .
\end{align}
To find an expression for $\Psi '$, we can take the derivative of Eq.~\eqref{eq:psieq} to get,
\begin{align}
\Psi' &= \Phi '+\frac{\ovf'+\ovU'}{\left(1+\ovf+\ovU\right)^{2}}\left[\frac{12 \pi G \, a^{2}}{k^{2}}\left(\bar{\rho}+\bar{p}\right)\sigma+\delta f+\delta U \right]
\nn\\ 
& \hspace{3cm} -\frac{1}{1+\ovf+\ovU}\left[\frac{24\pi G \, a^2 }{k^{2}}\mathcal{H}\left(\bar{\rho}+\bar{p}\right)\sigma+\frac{12\pi G \, a^{2}}{k^{2}}\left[\left(\bar{\rho}+\bar{p} \right)\sigma\right]'+\delta f '+\delta U' \right] \, .
\end{align}
The full evolution system of linear cosmological perturbation equations is closed with the addition of the energy-momentum conservation equations of each individual matter species considered. We start the evolution in deep radiation era and provide vanishing initial conditions for the linear perturbations of the auxiliary fields $\delta U, \delta X$, in agreement with our ``minimal'' choice for the boundary conditions. \\

\subsubsection{Tensor Sector}

The evolution equations for the linear cosmological perturbations of the traceless-transverse part of the spatial $3$--metric are given by (see also Ref.~\cite{Koivisto2008b}),
\begin{align}
&  h''_{ij} + 2 \hc \bigg( 1  - \frac{1}{2 \hc} \partial_\tau \log \big( G_{\rm eff, gw}(\tau)/G \big)  \bigg) h'_{ij}  + k^2 h_{ij}  = 16 \pi G_{\rm eff, gw} (\tau) \, a^2 \, \pi_{ij}  \, .\label{eq:GWprop}
\end{align}
where we have defined,
\begin{align}
\label{rescaledG}
G_{\rm eff, gw}(\tau)/G \equiv \bigg( 1 + f \big( \ovX(\tau) \big) + \ovU(\tau)  \bigg)^{-1} \, .
\end{align}
Several comments are in order at that point. First, we observe that within the DW model, the propagation equations for GWs are modified in their Hubble friction term (i.e., the coefficient of $h'_A$) as well as in their coupling to matter with respect to the one described in GR. More precisely, the extra quantities modifying such a behaviour identify themselves in terms of the ``Newton constant'' for GWs $G_{\rm eff, gw}(\tau)/G$, which is in fact the $|k| \rightarrow + \infty$ asymptotic behaviour of the effective Newton's constant $G_{\rm eff}(z,k)$ within the DW model. 
The latter is related to the modified growth and lensing features as compared to the same observables of the $\lcdm$ model, as we will discuss in detail in Secs.~\ref{sec:pheno} and~\ref{sec:MCMC}. Such a structure including $G_{\rm eff, gw}(\tau)/G$, is a quite typical fact in modified gravity theories, as is witnessed by the appearance of the same structure in the RR nonlocal gravity model of Ref.~\cite{Maggiore:2014sia} (see Refs.~\cite{Belgacem:2017ihm,Belgacem:2018lbp}) and generically in Horndeski models, see e.g. Refs.~\cite{Saltas:2014dha,Linder:2018jil}.

Second, and as a consequence, this structure in particular implies that the GWs propagate at the speed of light, as we will see below. Moreover, the fact that the Hubble friction is altered modifies the amplitude of the GWs as they propagate though spacetime, e.g. from inspiralling binaries of compact objects to observers\footnote{Actually, the amplitude of the GWs is rescaled as $(1 + f(0))^{-1}$ even in asymptotically flat space, in which $\ovX = 0 =\ovU$, as noted in Ref. \cite{Chu:2018mld}.}. 
Noticing that, on a cosmological background in GR, the GW amplitude is inversely proportional to the luminosity distance for electromagnetic sources $1/D_L(z)$, the accurate knowledge of the amplitude of their (polarised) strain and of their redshift, obtained for instance from an electromagnetic counterpart (and modulo systematic proportionality factors), therefore allows one to build up a Hubble diagram for compact binaries emitting GWs, making them ``standard sirens'' \cite{Schutz:2001re,Holz:2005df}. Modifying the friction term in Eq.~\eqref{eq:GWprop}, makes the relations between GWs amplitude and luminosity distance different, forming a luminosity distance for GWs $D_L^{\rm gw}(z)$, whose Hubble diagram therefore differs from the electromagnetic one. When the propagation equations for GWs are modified such as in Eq.~\eqref{eq:GWprop}, both notions are related as \cite{Belgacem:2017ihm,Amendola:2017ovw,Linder:2018jil}, 
\begin{align}
D_L^{\rm gw}(z) \equiv D_L^{\rm em}(z) \times \sqrt{\frac{G_{\rm eff, gw}(z)}{G_{\rm eff, gw}(z=0)}} \, .
\end{align}

As we will discuss in more details in Sec.~\ref{sec:devGRtens}, the ratio between the Hubble diagrams for electromagnetic and gravitational signals provides a powerful way for testing deviations to GR with future third generation GW interferometers such as the \textit{Einstein Telescope} \cite{ET} or the \textit{Cosmic Explorer} \cite{CE}. 

We now display the result of the reconstruction procedure conduced  in Ref.~\cite{Deffayet2009}.

\subsection{Distortion Function}\label{sec:funcf}

\begin{figure}[h!]
\begin{center}
\vspace{-0.15cm}
\includegraphics[width=0.495\columnwidth]{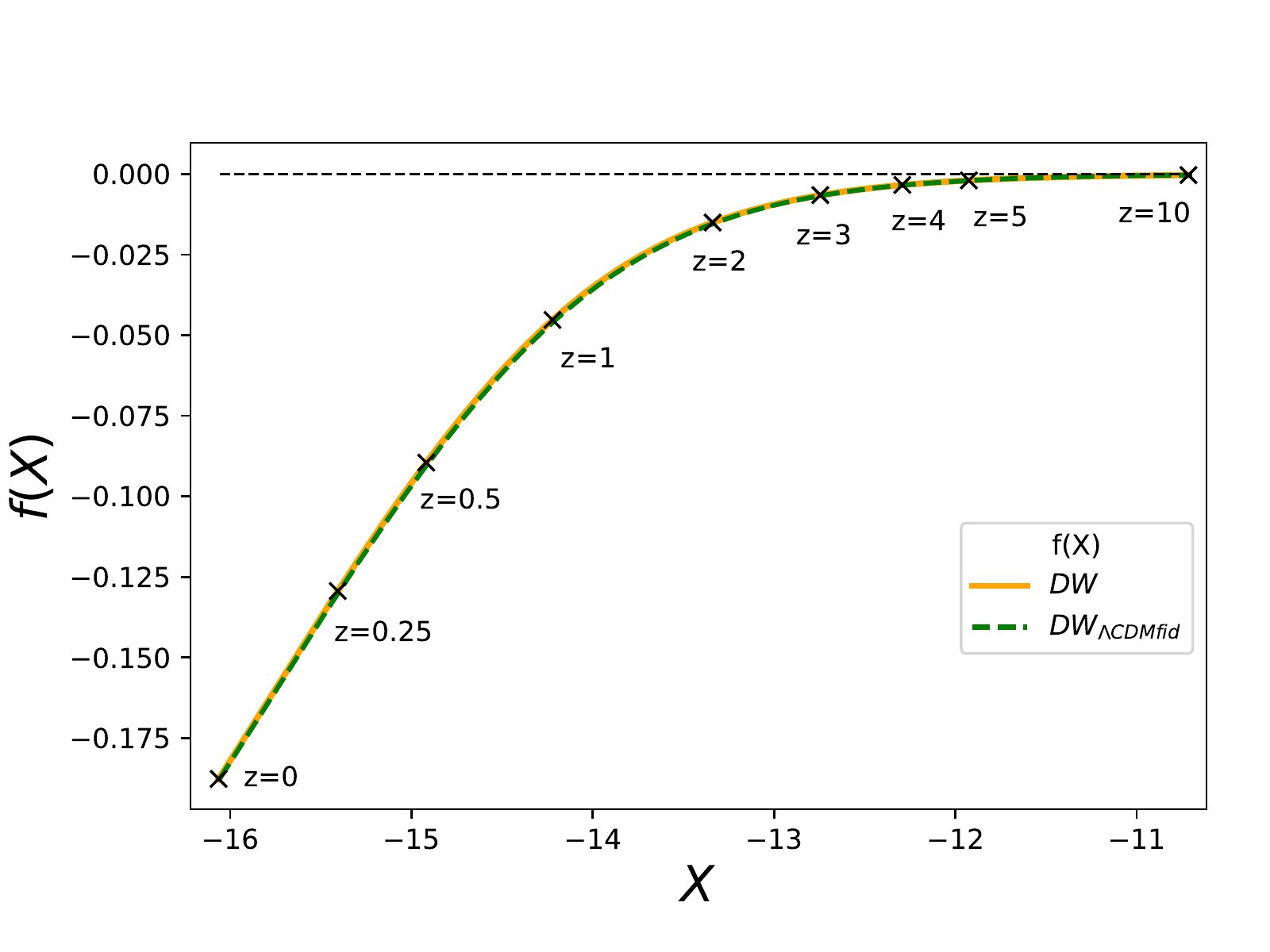} 
\includegraphics[width=0.495\columnwidth]{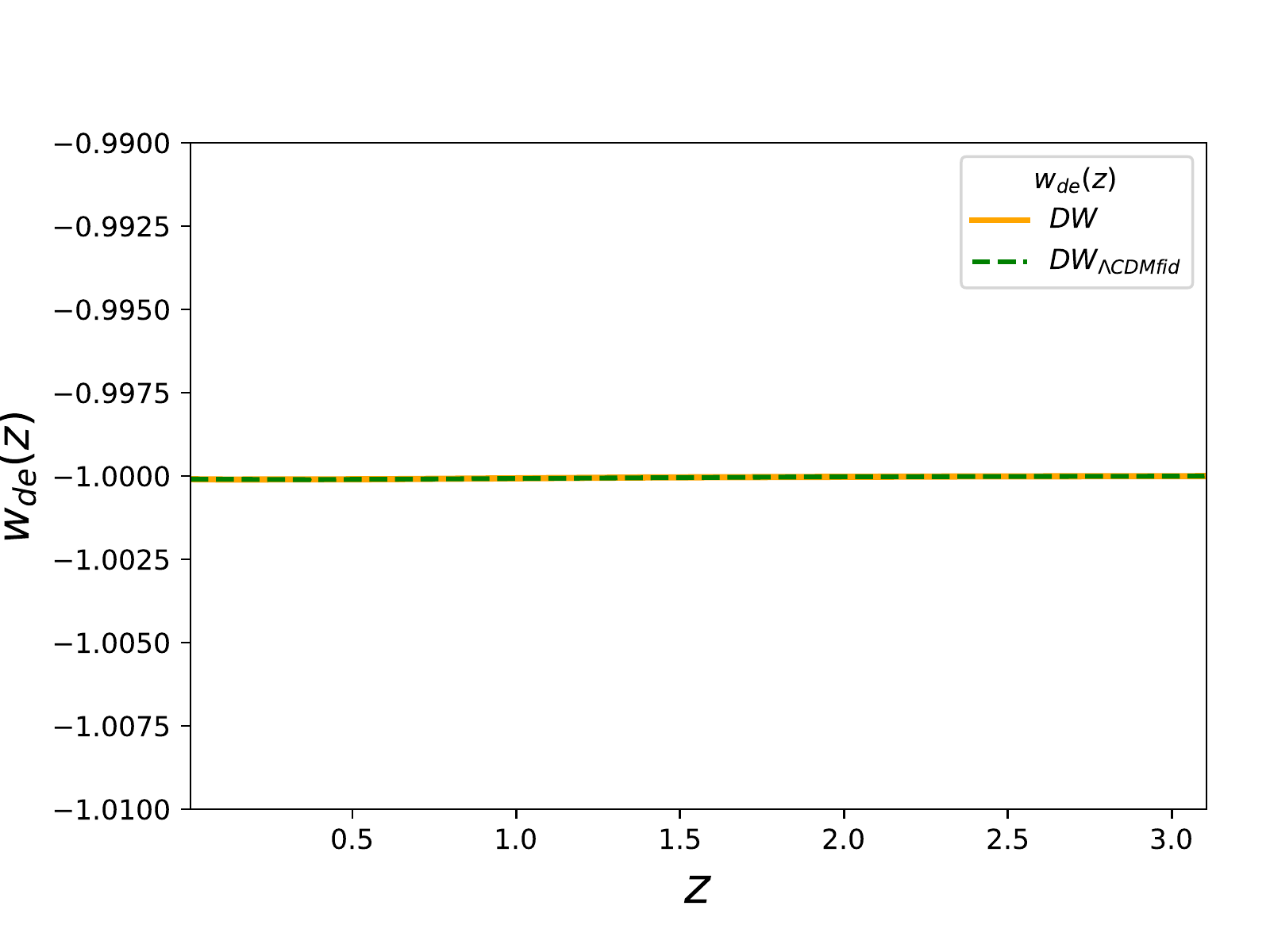}
\vspace{-0.5cm}
\caption{\textit{Left panel:} the distortion function $f$ as a function of $\ovX$. The black crosses indicate the functions corresponding values for different values of $z$. \textit{Right panel:} effective dark energy equation of state. In both plots, the orange solid lines shows the prediction from DW on its bestfit to the CMB+SNIa+RSD data we describe in Sec.~\ref{sec:data}, while the green dashed line shows the prediction from DW on the $\lcdm$ best fitting parameter values to the same data, a cosmological model that we quote $\dwlcdm$ in the following.}
\label{fig:UfX}
\vspace{-0.5cm}
\end{center}
\end{figure}

Following Ref.~\cite{Deffayet2009}, for reproducing the $\Lambda$CDM expansion history within the DW model, for non--trivial distortion function $f\left(\ovX\right)$, the latter is expressed in terms of the reduced $\lcdm$ Hubble expansion rate,
\begin{equation}
h^2_\Lambda(\zeta) \equiv H_{\rm \Lambda}^{2}\left(\zeta\right)/H_0^2 = \big[ \Omega_{\Lambda}+\Omega(\zeta) \big] \, .
\end{equation}
where $\zeta \equiv 1/a = 1+z$ and $z$ is the redshift, $H_0$ is the Hubble constant today and the total energy density fraction of the different matter species reads $\Omega(\zeta) = \sum_i \Omega_{i}(\zeta) \equiv \sum_i 8 \pi G \, \bar{\rho}_i(\zeta) / (3 H_0^2)$, and for the dark energy $\Omega_{\Lambda} \equiv \Lambda / (3 H_0^2)$. The reconstruction procedure leads to,
\begin{align}\label{ffunc}
f\left(\zeta\right)=&-2\int_{\zeta}^{\infty} \dd\zeta_{1} \,  \zeta_{1} \phi\left(\zeta_{1}\right)-6\Omega_{\Lambda}\int^{\infty}_{\zeta} \dd \zeta_{1} \, \frac{\zeta_{1}^{2}}{\hcl\left(\zeta_{1}\right)I\left(\zeta_{1}\right)}\int_{\zeta_{1}}^{\infty}\dd \zeta_{2} \, \frac{I\left(\zeta_{2}\right)}{\hcl\left(\zeta_{2}\right)\zeta_{2}^{4}} \nonumber \\
&\hspace{6cm}+2\int_{\zeta}^{\infty} \dd \zeta_{1} \frac{\zeta_{1}^{2}}{\hcl\left(\zeta_{1}\right)I\left(\zeta_{1}\right)}\int_{\zeta_{1}}^{\infty} \dd \zeta_{2} \, \frac{r\left(\zeta_{2}\right)\phi\left(\zeta_{2}\right)}{\zeta_{2}^{5}} \, ,
\end{align}
with
\begin{equation}
r(\zeta) \equiv \bar{R}/ H_{0}^{2} = 6\left( \frac{\hcl'}{a H_0} + 2 \hcl^2 \right) \, ,
\end{equation}
and
\begin{equation}
\phi \left(\zeta\right)=-6 \Omega_{\Lambda}\int^{\infty}_{\zeta} \dd \zeta_{1} \, \frac{1}{\hcl\left(\zeta_{1}\right)}\int_{\zeta_{1}}^{\infty}\dd \zeta_{2} \, \frac{1}{\hcl\left(\zeta_{2}\right)\zeta_{2}^{4}} \, , \qquad \qquad I\left(\zeta\right)=\int_{\zeta}^{\infty}\dd \zeta_{1} \, \frac{r\left(\zeta_{1}\right)}{\zeta_{1}^{4}\hcl\left(\zeta_{1}\right)} \, .
\end{equation}
In order to solve the background system of equations, one can numerically integrate Eq.~\eqref{ffunc} and, at any timestep, obtain $f$ as a function of $\zeta$. One can then compose it with the inverse function of,
\begin{equation}
\ovX\left(\zeta\right)=-\int_{\zeta}^{\infty}\frac{\dd \zeta_{1} \, \zeta_{1}^{2}}{\hcl \left(\zeta_{1}\right)}I\left(\zeta_{1}\right) \, ,
\end{equation}
to obtain $f(\ovX)$. In the left panel of Fig.~\ref{fig:UfX}, we show two reconstructions of the function $f(\ovX)$ for different sets of cosmological parameters. We also display the points $(f,X)(z_i)$ as a function of several redshift values  $z_i$, marked by black crosses. The $\lcdm$ expansion history is reproduced within the DW model via the effective dark energy it describes, which features a constant equation of state $w_{\textsf{de}}=-1$ as shown in the right panel of Fig.~\ref{fig:UfX}. Hence, we see that the DW modification to gravity $\sim R f(\Box^{-1}R)$ can behave exactly the same way as a cosmological constant, provided the above results of the reconstruction procedure worked out  in Ref.~\cite{Deffayet2009} are used. As we will see in the next section, the fact that the background expansion history is the same as $\lcdm$, does not imply that the linear cosmological perturbations described by the DW model are necessarily the same than in $\lcdm$. Later, we will also see that this fact makes the DW models attractive in the light of current high precision cosmological data.  

%%%%%%%%%%%%%%%%%%%%%%%%%%%%%%%%%%%%
\section{Cosmological Phenomenology}\label{sec:pheno}
%%%%%%%%%%%%%%%%%%%%%%%%%%%%%%%%%%%%

In this section, we integrate the background and linear perturbation systems using a modified version of the linear Boltzmann-Einstein solver CLASS \cite{CLASS}. To do so, we fix both models' cosmological setting to the so--called \textit{Planck} baseline described in detail in Refs.~\cite{Planck_2013_CP,Planck2015CP} which, in particular, is parametrized by six cosmological parameters (see Sec.~\ref{sec:MCMC} for more details). We first focus on deviations to GR in the scalar sector and then on deviations in the tensor sector.

\subsection{Scalar Indicators from GR--deviations}\label{sec:devGRsca}

From the evolution equations presented in Sec.~\ref{sec:eqlin}, we integrate the scalar linear cosmological perturbations in the DW model. We use adiabatic initial conditions from a Gaussian random field with a slightly red tilted flat power spectrum.
To evaluate the extent to which the DW model deviates from GR within the scalar sector, it is convenient to introduce the following indicative functions \cite{Zhang:2007nk,Amendola2007,Daniel2010}\footnote{For these four functions, we work on standard $\lcdm$ best fitting cosmological parameter values inferred from observational constraints given the CMB+SNIa+RSD joined data presented in Sec.~\ref{sec:MCMC} (see also Table~\ref{tab:partable}).},
\begin{align}
\eta(z,k) &\equiv \frac{\Psi}{\Phi} \, , \label{eq:eta} \\
G_{\rm eff}/G(z,k) &\equiv \frac{k^2 \Phi}{4 \pi G \,  \bar{\rho} \, a^2 \delta} \, , \label{eq:GeffG}\\
\Psi &\equiv \big(1+\mu(z,k) \big) \Psi_{\rm \Lambda CDM} \, , \label{eq:mu}\\
\big( \Psi + \Phi \big) &\equiv \big( 1+\Sigma(z,k) \big) \big( \Psi_{\rm \Lambda CDM} + \Phi_{\rm \Lambda CDM} \big) \, . \label{eq:sig}
\end{align}
\begin{figure}[h!]
\begin{center}
\vspace{-0.25cm}
\includegraphics[width=0.495\columnwidth]{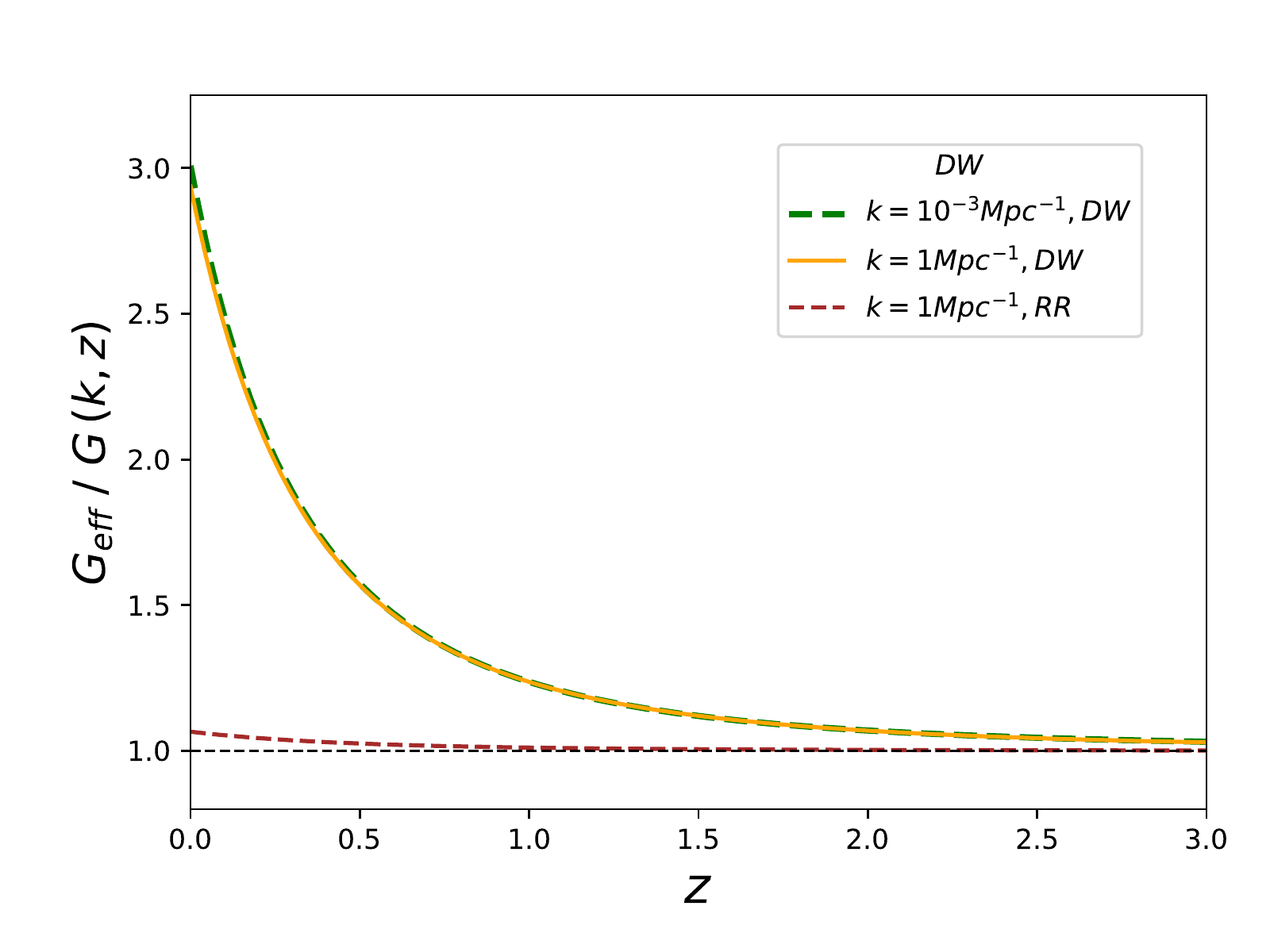}
\includegraphics[width=0.495\columnwidth]{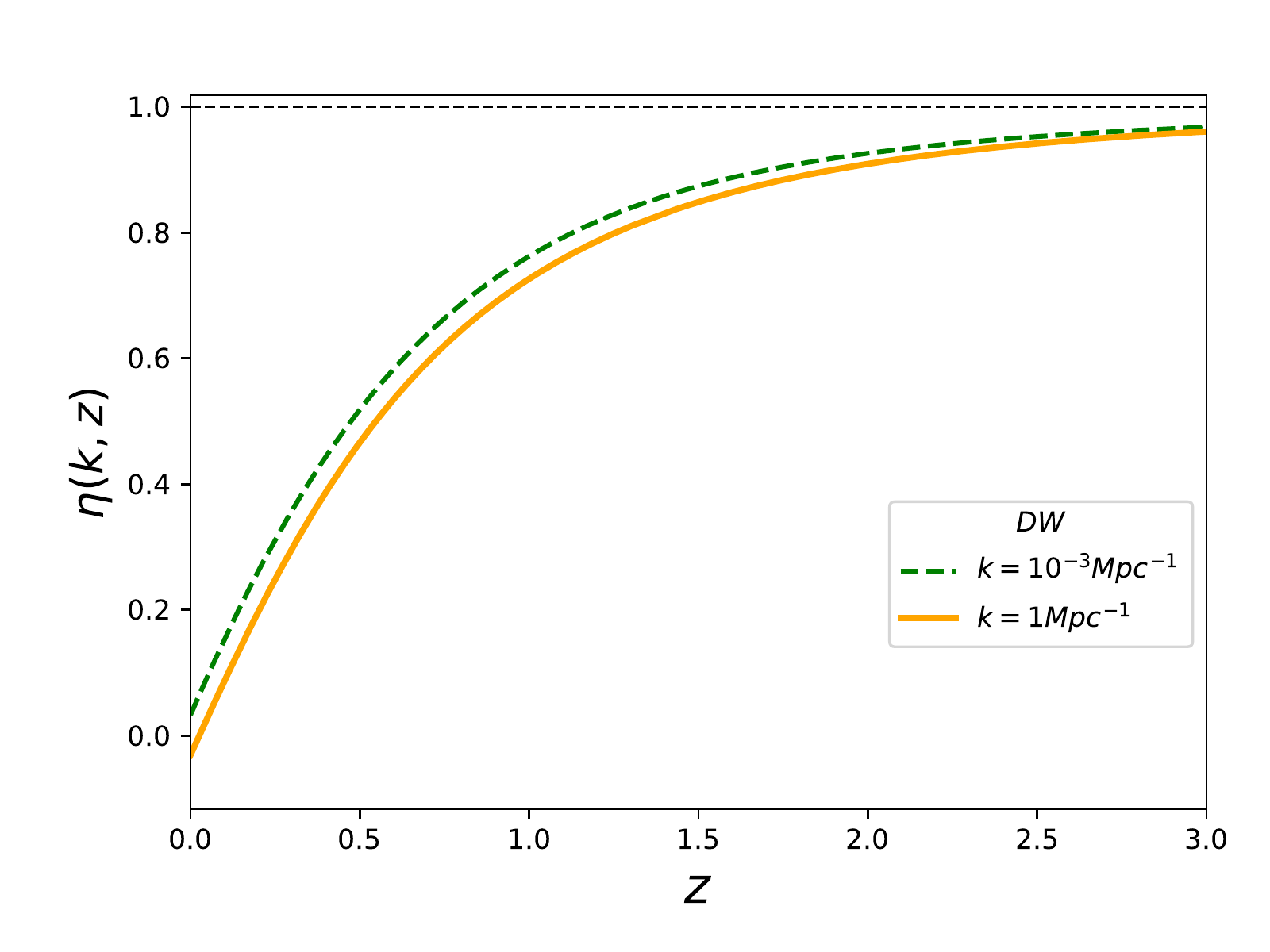}
\vspace{-0.75cm}
\caption{The effective Newton constant $G_{\rm eff}$ (left panel) and the inverse of the gravitational slip $1/\eta$ for $k=10^{-3} {\, \rm Mpc^{-1}}$ (green dashed) and for $k=1 {\, \rm Mpc^{-1}}$ (solid orange).}
\label{fig:Geffeta}
\vspace{-0.25cm}
\end{center}
\end{figure}
where $\delta$ is the linear gauge invariant matter density contrast of total matter, $\eta$ is the gravitational slip and $G_{\rm eff}$ is the effective Newton constant.
The quantity $\mu$ is the deviation of the gravitational potential $\Psi$ in the DW model with respect to the one in $\lcdm$, whereas $\Sigma$ measures deviations in the lensing (Weyl) potential. The quantity $\mu$ therefore relates to the modification of the motion of non-relativistic matter and is therefore probed through clustering properties of structures (growth) while $\Sigma$ relates to the motion of relativistic particles (e.g. light) and is probed through WL. Another set of alternative indicators to $\{ \eta, \mu, \Sigma \}$ is given by,
\begin{align}
\eta_p(z,k) &\equiv \eta^{-1}(z,k) \, , \label{eq:etap}\\
\mu_p(z,k) &\equiv - \frac{k^2 \Psi(z,k)}{4 \pi G \, a^2 \bar{\rho} \delta}=-\eta G_{\rm eff}/G(z,k) \, , \label{eq:mup}\\
\Sigma_p(z,k) &\equiv - \frac{k^2 (\Psi + \Phi)}{8 \pi G \, a^2 \bar{\rho} \delta} \, ,\label{eq:sigp}
\end{align}
where we denote with a subscript $_{P}$ the quantities used  in Ref.~\cite{Planck2015DE}. In that case, all the terms on the right hand sides of the three above equations are evaluated within a given model and no comparisons between the true and alternative hypothesis are made. 
\begin{figure}[t]
\begin{center}
\vspace{-0.5cm}
\includegraphics[width=0.45\columnwidth]{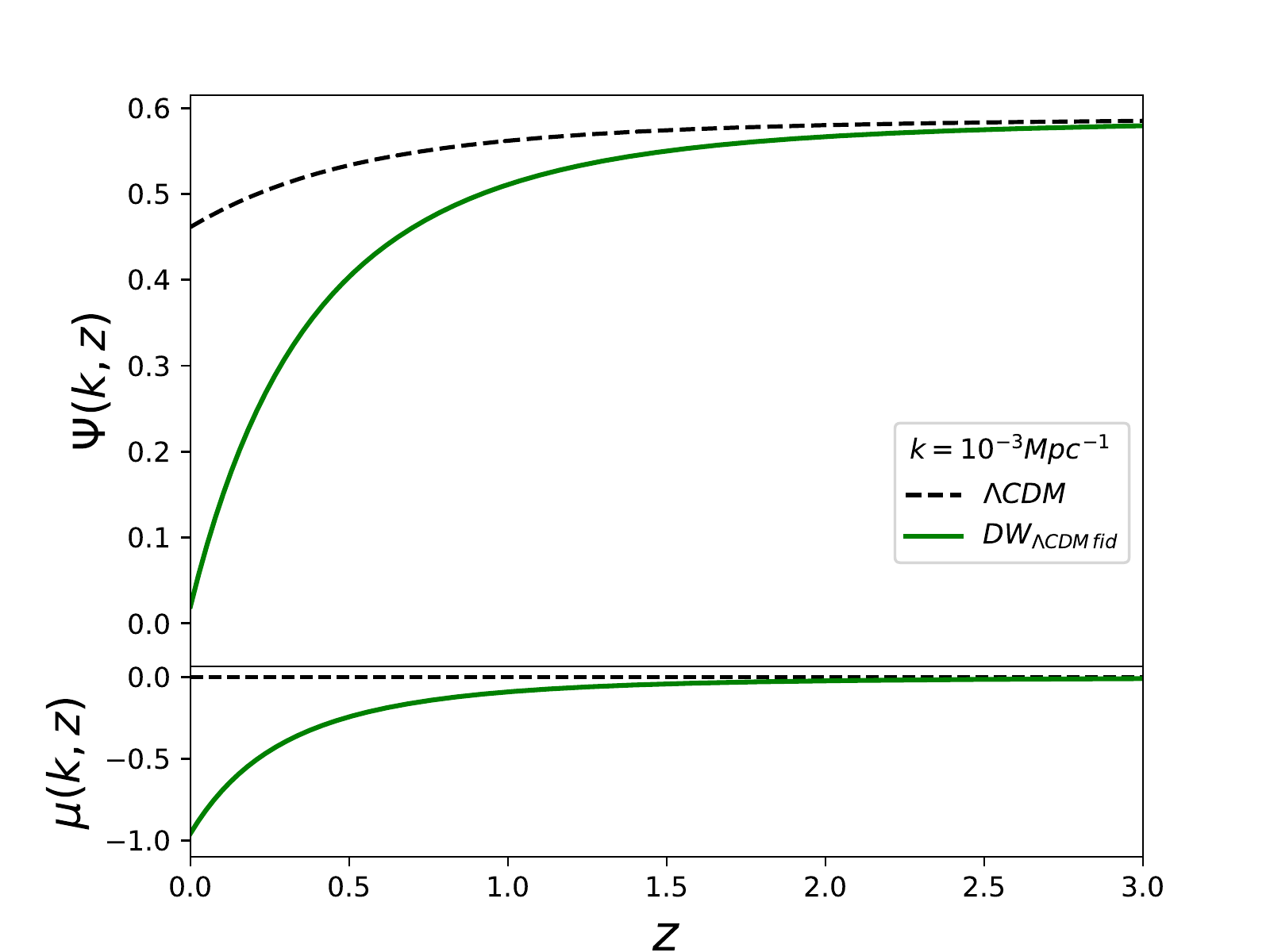}
\includegraphics[width=0.45\columnwidth]{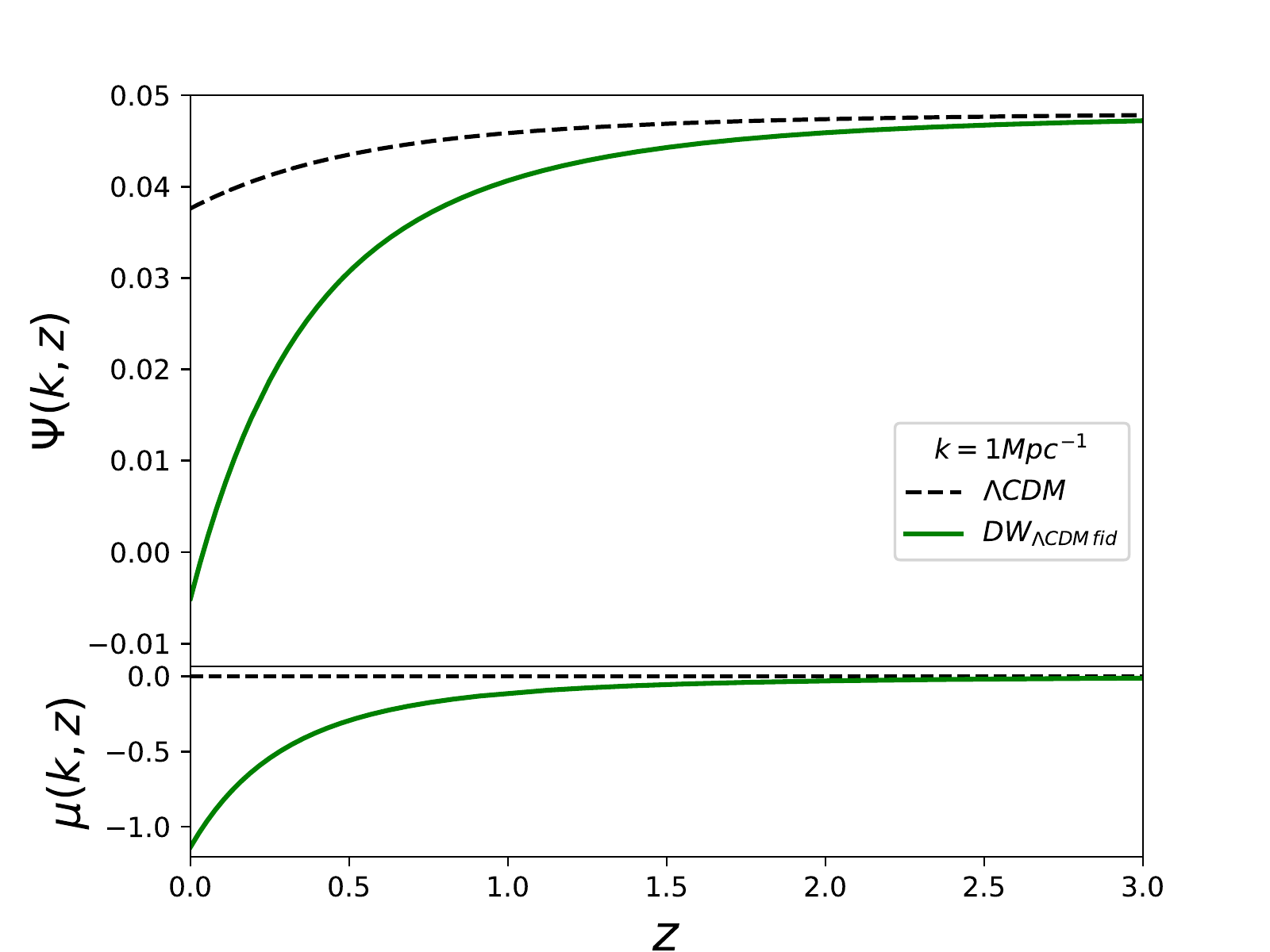}
\vspace{-0.15cm}
\caption{The gravitational potential $\Psi$ in $\lcdm$ (black dashed) and in $\dwlcdm$ (green solid) for $k=10^{-3} {\, \rm Mpc^{-1}}$ (left upper panel) and for $k=1 {\, \rm Mpc^{-1}}$ (right upper panel). The lower panels reproduces the indicator $\mu$ as a function of redshift for both chosen length scales, respectively.}
\label{fig:mu}
\vspace{-0.5cm}
\end{center}
\end{figure}
The latter quantities probe the response of the gravitational potential $\Psi$, and of the lensing potential $\Psi + \Phi$, to the total distribution of matter fluctuations within a given gravity model. We display the results of both parametrisations for completeness. 
\begin{figure}[h!]
\centering
\vspace{-0.25cm}
\includegraphics[scale=0.6]{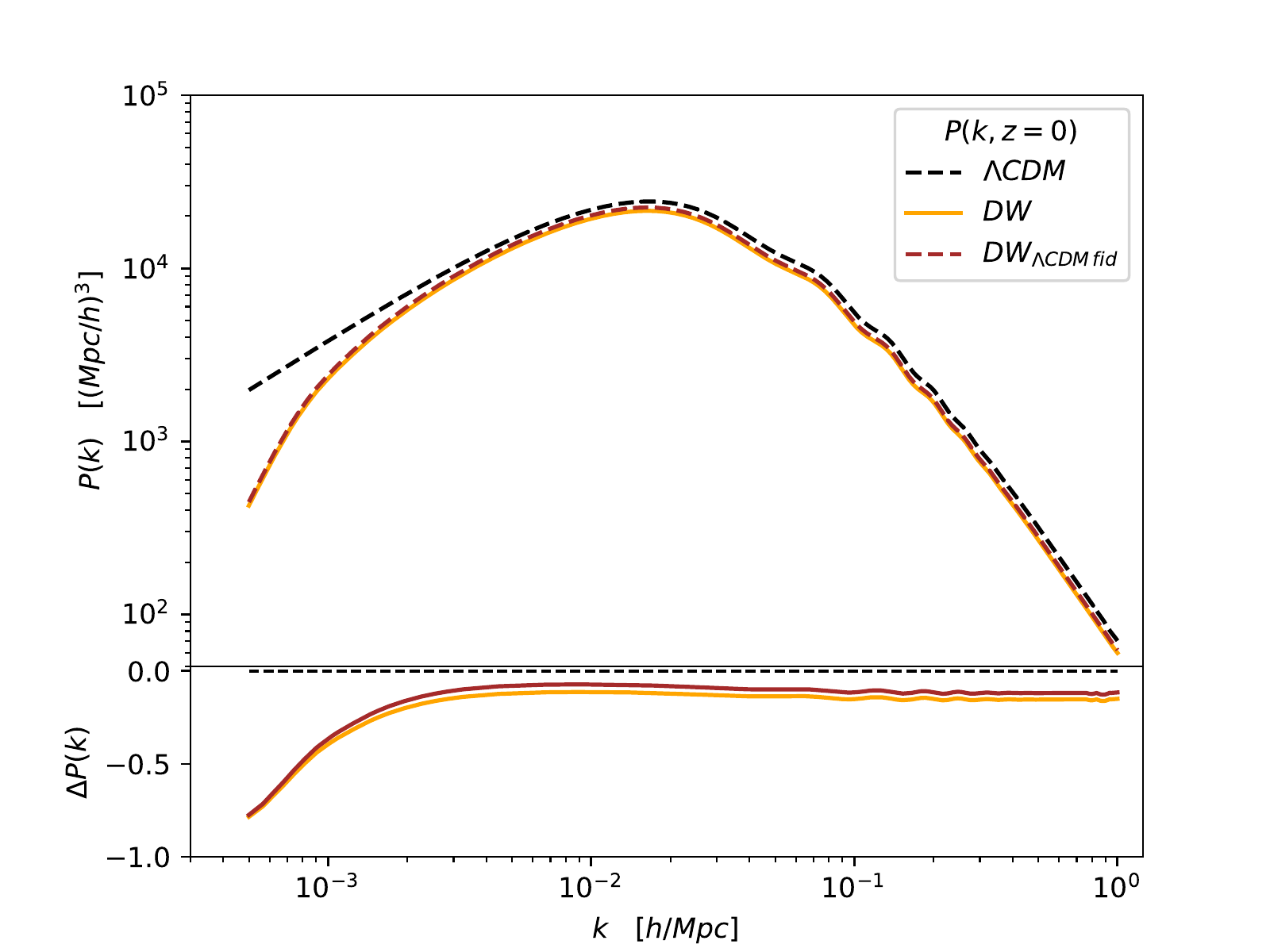}
\vspace{-0.15cm}
\caption{\label{fig:mPk} \textit{Upper panel: }Gauge invariant linear matter power spectrum at $z=0$ computed for the $\Lambda$CDM (black dashed), the DW (orange solid) on their respective best fit to \textit{Planck} CMB data and for the $\dwlcdm$ model (brown dashed). \textit{Lower panel:} corresponding relative differences.}\vspace{-0.5cm}
\end{figure}

The quantities in Eqs.~\eqref{eq:eta}--\eqref{eq:sig} are evaluated on the same cosmological parameter values, the best fitting of standard $\lcdm$ to CMB+SNIa+RSD data described in Sec.~\ref{sec:data}, so that the predicted background cosmologies of both models are similar and the relevant differences lie in their linear cosmological perturbations. In the scalar sector of the theory, these are conveniently parametrised by pairs drawn from the set $\{ \eta, G_{\rm eff}, \mu, \Sigma \}$, and very similarly in the tensor sector as will be discussed below.
\begin{figure}[h!]
\begin{center}
\vspace{-0cm}
\includegraphics[width=0.45\columnwidth]{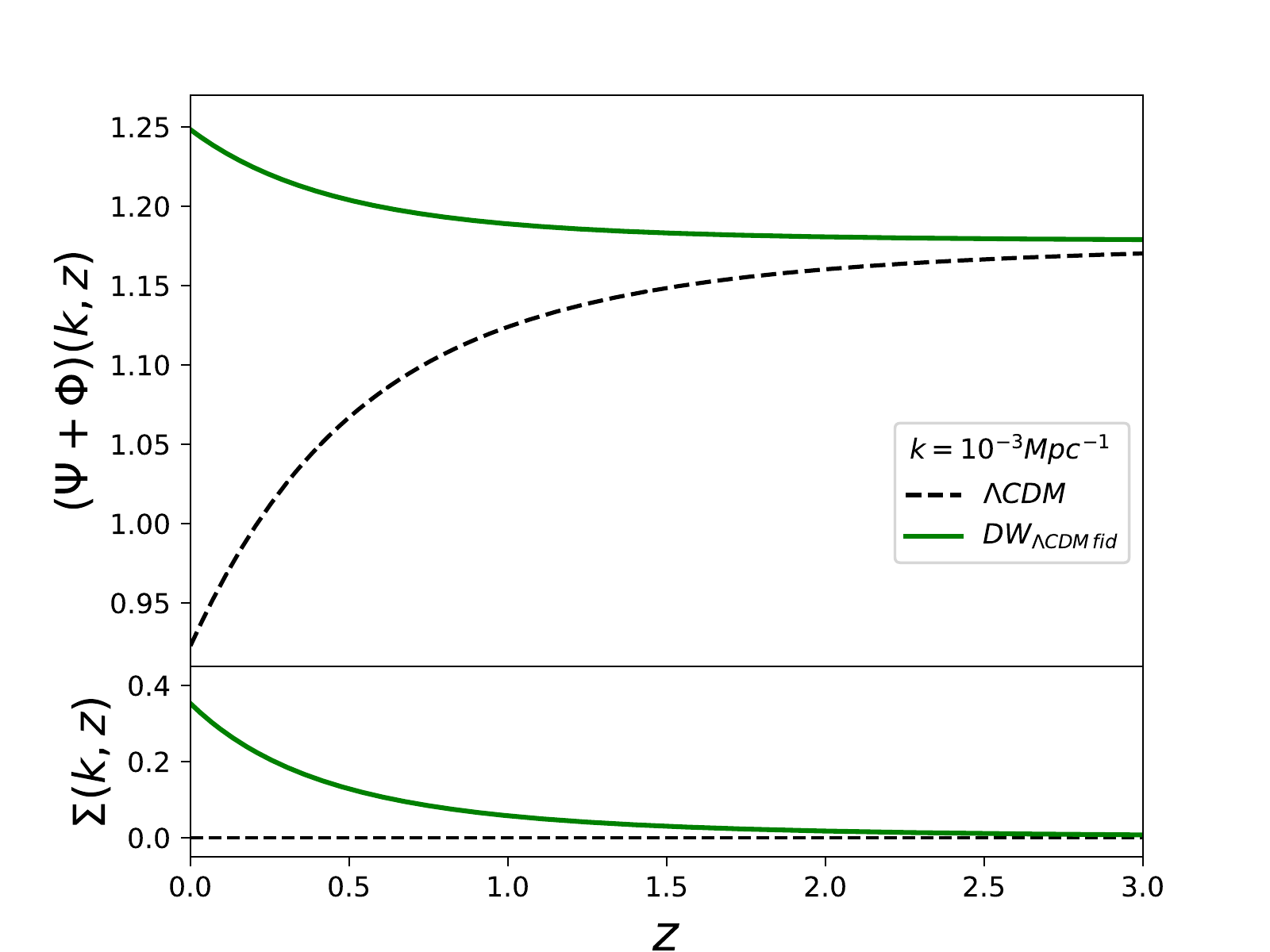}
\includegraphics[width=0.45\columnwidth]{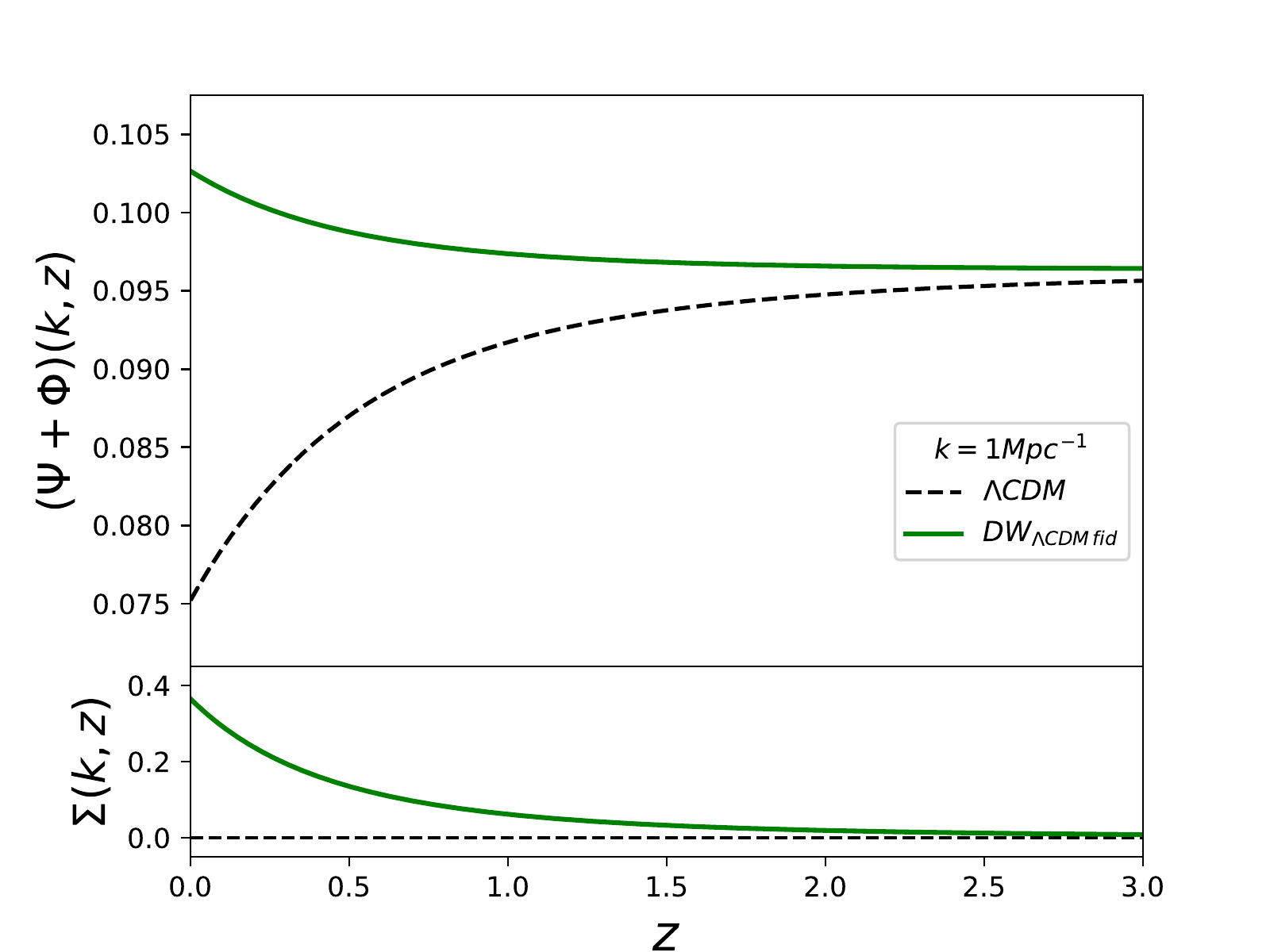}
\vspace{-0.15cm}
\caption{The lensing potential $\Psi - \Phi$ in $\lcdm$ (black dashed) and in $\dwlcdm$ (green solid) for $k=10^{-3} {\, \rm Mpc^{-1}}$ (left upper panel) and for $k=1 {\, \rm Mpc^{-1}}$ (right upper panel). The lower panels show the ratio between both minus one, which reproduces $\Sigma$ as a function of redshift for both chosen wave numbers.}
\label{fig:sig}
\vspace{-0.5cm}
\end{center}
\end{figure} 
Such a setting is convenient to express ``how far'' the alternative model deviates from GR for given features probed by cosmological surveys, such as in the CMB, SNIa or distribution of galaxies. The second one, formed by pairs drawn from the set $\{ \eta_P, G_{\rm eff}, \mu_P, \Sigma_P \}$, which is more model independent, is most conveniently used for forecasting \cite{Casas:2017eob} or constraining \cite{Planck2015DE} deviations from GR given future or current data respectively. 

\newpage 

In what follows, we work a posteriori and anticipate the observational constraints results provided CMB+SNIa+RSD data presented  in Sec.~\ref{sec:MCMC}.
When showing the quantities in Eqs.~\eqref{eq:eta}--\eqref{eq:sig}, we fix both models' cosmological parameters to the $\lcdm$ best fitting values given CMB+SNIa+RSD data of Sec.~\ref{sec:data} and we quote such DW cosmological model as $\dwlcdm$. The indicators in Eqs.~\eqref{eq:etap}--\eqref{eq:sigp} are displayed on the CMB+SNIa+RSD best fitting parameter values of each respective model.
The left panel of Fig.~\ref{fig:Geffeta} shows the effective Newton constant $G_{\rm eff}/G$, as a function of redshift at large and small scales. The function is the same for both scales and modifies the response of the gravitational potential $\Phi$ to the fluctuations of matter. In the case of the DW model, such a response is enhanced and can leave significant imprints in the predictions of galaxy clustering features. More precisely, given that the growth is mostly controlled by $\Psi$ (as is seen from the growth equation, see e.g. Eq.~4.45 of Ref.~\cite{Dirian:2014ara}) which relates to $\Phi$ through Eq.~\eqref{eq:psieq}, which in turn relates to $\eta$, the non--trivial behaviour of $\eta$ implies that deviations in clustering and lensing predictions are of opposite trends. Indeed, a trivial behaviour for $\eta$, i.e. $\eta(z,k) \approx 1$, implies that an enhanced $G_{\rm eff}(z)$ directly translates into an enhanced growth of structures probed by $\mu$, as well as a stronger lensing power probed by $\Sigma$ (examples of such models are provided  in Ref.~\cite{Dirian:2014ara}). However, in the case for the DW model, $\eta$ has a non-trivial behaviour as seen from the right panel of Fig.~\ref{fig:Geffeta}. The trend (this is not an exact limit) $\eta(z \rightarrow 0, k) \rightarrow 0$, reflects the fact that the anisotropic stress of the effective dark energy described by the DW model drives the linear perturbation of the potential $\Psi$ to small values at small redshifts. This shows that, although $G_{\rm eff}/G(z)$ is greater than unity in the DW model, the clustering of linear structures is lowered compared to the one described by $\lcdm$. This is induced by a non-trivial behaviour of the anisotropic stress associated with the effective dark energy described by the DW model.
\begin{figure}[h!]
\begin{center}
\vspace{-0.4cm}
\includegraphics[width=0.45\columnwidth]{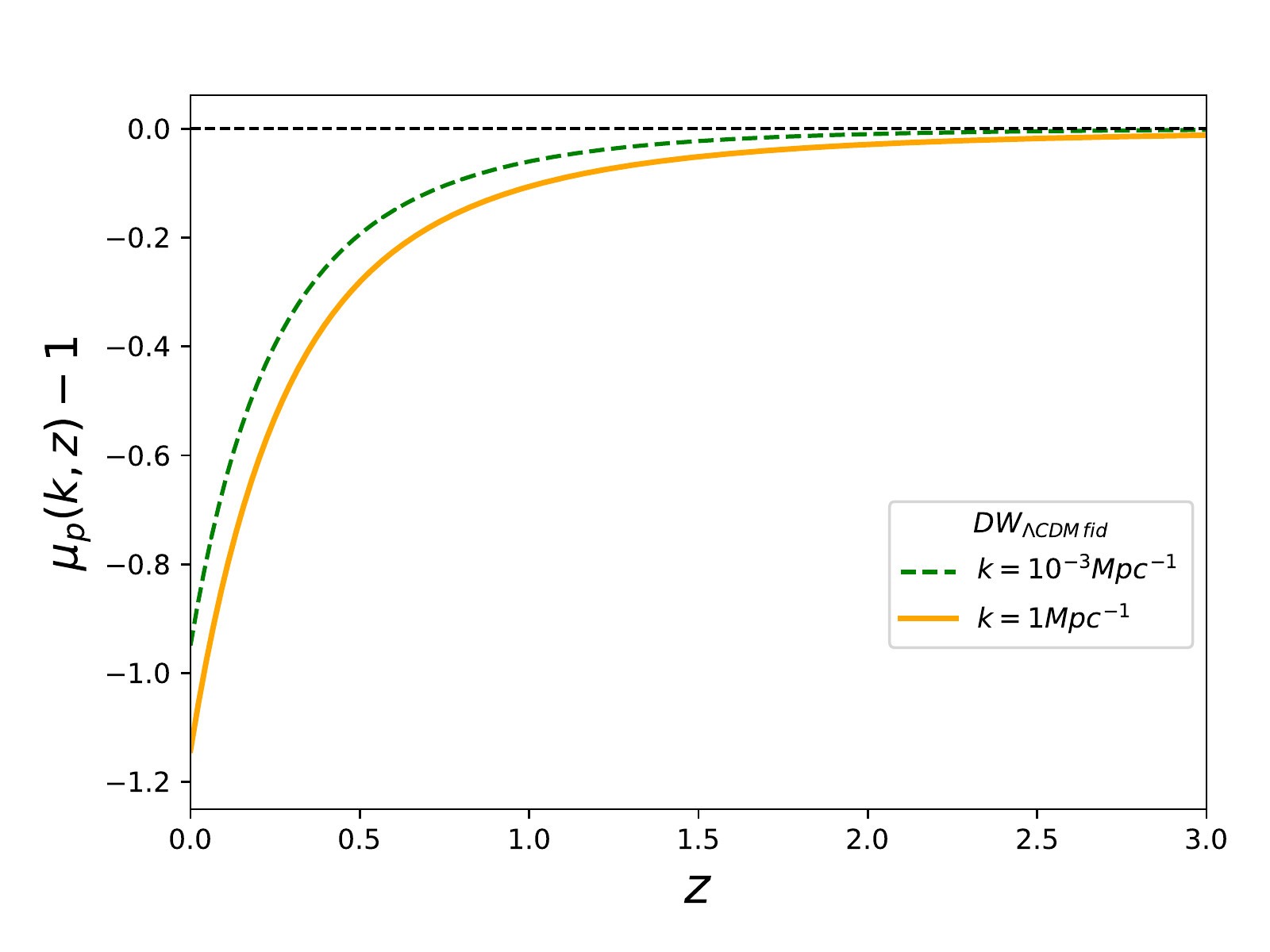}
\includegraphics[width=0.45\columnwidth]{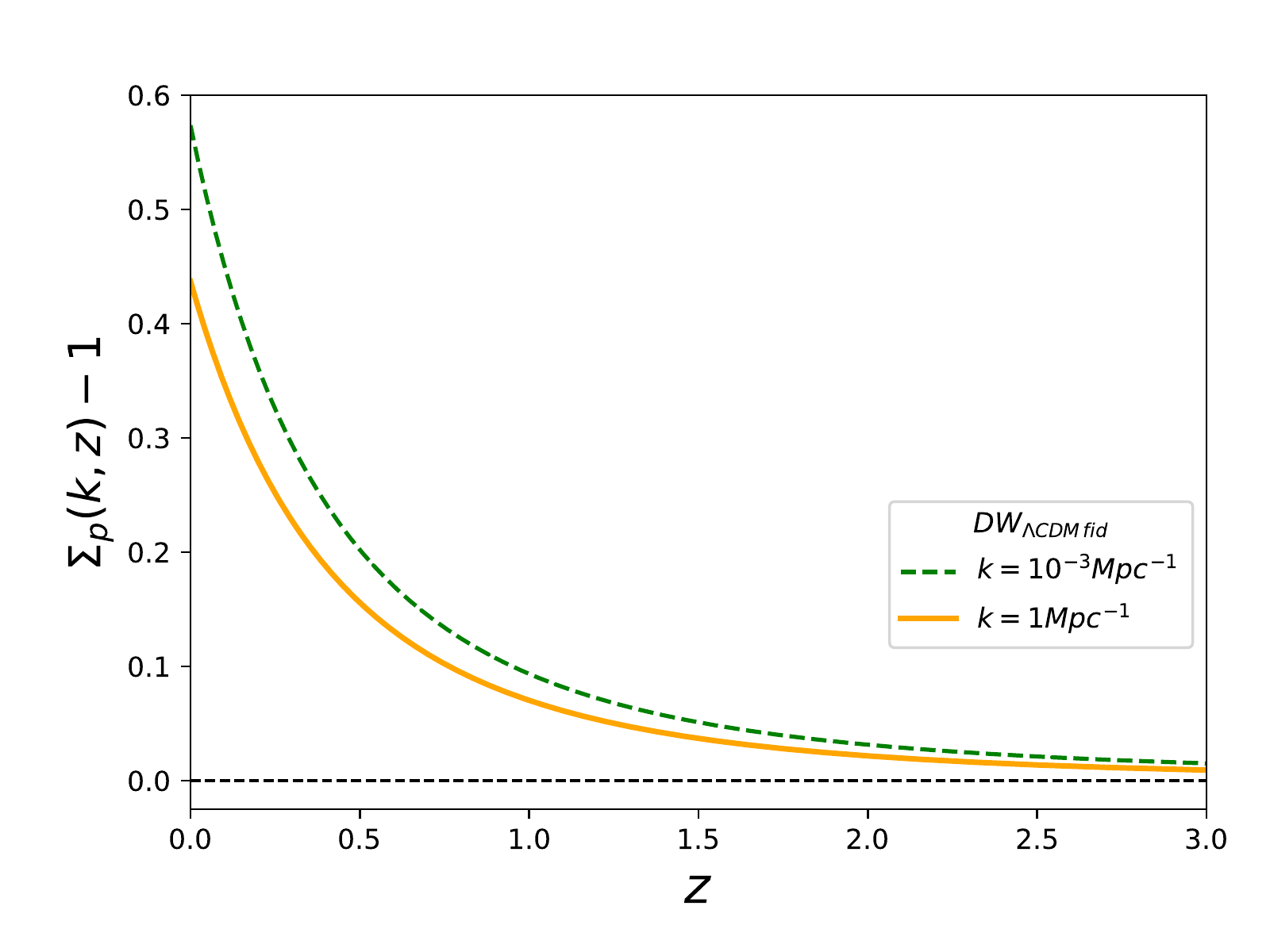}
\vspace{-0.5cm}
\caption{The quantities $\mu_P(z,k)$ (left panel) and $\Sigma_P(z,k)$ (right panel), in the $\dwlcdm$ model. These are defined according to the convention adopted in Ref.~\cite{Planck2015DE}. The cosmological parameters are chosen on the respective models' best fitting values to CMB+SNIa+RSD data. Both plots show these quantities for $k=10^{-3} {\, \rm Mpc^{-1}}$ (green dashed) and for $k=1 {\, \rm Mpc^{-1}}$ (orange solid).}
\label{fig:mup}
\vspace{-0.5cm}
\end{center}
\end{figure} 

This fact is illustrated in Fig.~\ref{fig:mu}, which shows the deviations of $\dwlcdm$ to $\lcdm$ in $\Psi(z)$. We see that the latter is lower  in the DW model at large and small scales, and the deviation increases at very late time. This has a significant impact on the growth of structures as is seen from Fig.~\ref{fig:mPk}, which shows the total matter power spectrum within the DW model as compared to the one described by $\lcdm$ (on their respective best fit values given CMB+SNIa+RSD data), or $\dwlcdm$. The power spectrum predicted by the $\dwlcdm$ model (i.e. DW on the same cosmological parameter values as $\lcdm$) is lower than the one given by $\lcdm$ by a constant factor of $10 \%$ down to scales of about $k \approx 10^{-2} \,{\rm Mpc^{-1}}$, below which the deviation of the DW model to GR drops further down. 
Turning to the lensing potential shown in Fig.~\ref{fig:sig}, we see that a non-negligible anisotropic stress affects the lensing considerably. The latter is pushed in the opposite direction as compared to the deviation in the growth.
While a lower growth of structures would intuitively involve a deficit in the lensing response as well, or vice versa (such as in the RR nonlocal gravity model studied in Ref.~\cite{Dirian:2014ara}), we can see that in the DW model the lensing potential is in fact enhanced by a few tens of percent at late time as compared to the one in $\lcdm$. 
The behaviour of the indicators $\mu$ and $\Sigma$ can be compared with the ones defined  in Eqs.~\eqref{eq:etap}--\eqref{eq:sigp} and which are shown in Fig.~\ref{fig:mup}. We can see that the scale dependence is more pronounced for the $_P$--quantities, but the results remain qualitatively the same, up to a few percents. The similarity of both parametrisations partly ties to the fact that the cosmological background in the DW model is the same as the one of $\lcdm$. In Refs.~\cite{Planck2015DE,Aghanim:2018eyx}, the authors inferred constraints on the present time values $\mu_0 \equiv \mu(z=0)$ and $\Sigma_0 \equiv \Sigma(z=0)$ (see their Fig.~15) and their possible scale dependence, through the use of phenomenological functions together with complementary cosmological data. In the case of the $\dwlcdm$ model, the present time values correspond to $(\mu_0, \Sigma_0) = ( -0.95, 0.57 )$ for $k=10^{-3} \,{\rm Mpc^{-1}}$, and therefore lie into the ``sweet quadrant'' for galaxy WL and galaxy clustering data such as RSD. \\

\newpage

\paragraph*{\hspace{-0.35cm}\textbf{Solar System Constraints.}} One of the possible drawbacks that arose in nonlocally modified gravity theories is the potential remaining of a FLRW background--time dependence in the small scale limit of the Newton constant, 
\begin{align}
G_{\rm eff}(z,k \gg 1) ~ \simeq ~ G_{\rm eff}(z) ~ \neq ~ G \, ,\label{eq:limGeff}
\end{align}
that is, a lack of screening mechanism that spoils the predictions of the theory at solar system scales.
In effects, as originally noticed in Ref.~\cite{2014JCAP...09..031B}, such a residual time-dependence can expose these models to dangerous conflicts with Lunar Laser Ranging experiments (LLR), that put bounds on the time variation of the Newton constant such as, $\dot{G}_{\rm eff}/G = (4\pm9) \, \times \, 10^{-13} \mathrm{yr^{-1}}$ \cite{Williams2004}.
An example of such a model is provided by the RR model introduced in Ref.~\cite{Maggiore:2015rma}, where the original Newton constant $G$ remains multiplied by a term depending on FLRW background quantities. In the case of the DW model, we see that, from the variation of the action Eq.~\eqref{eq:DWaction} evaluated in cosmological perturbation theory, the small scale asymptotics of the effective Newton constant is a background dependent function too (see also the discussion of Ref.~\cite{2014JCAP...09..031B}), 
\begin{align}
G_{\rm eff} / G (z,k \gg 1) = \bigg( 1 + f(\ovX) + \frac{1}{\osq}( \bar{R} f_{,}) \bigg)^{-1} \, ,
\label{Geff-bigk}
\end{align}
and its asymptotic behaviour on small scales is basically the one seen in the left panel of Fig.~\ref{fig:Geffeta}, on a FLRW cosmological background. 
In Ref.~\cite{DW2013} (see also Ref.~\cite{Woodard:2014iga}, for similar views), the authors argue that inside bound objects the auxiliary field $X = \Box^{-1}R$ is positive, whereas it is negative in cosmology (see left panel of Fig.~\ref{fig:UfX}). Then, their point is that, as one is free to choose the distortion function, one can set it so that it vanishes for positive values of $X$, i.e. $f(X) \sim \theta(-X)$, where $\theta$ is the Heaviside step function. Hence in that case, a ``perfect'' screening mechanism makes the DW model reproduce the well established predictions of GR on solar system scales.

However, as explicitly outlined in Ref.~\cite{Belgacem:2018wtb}, the value of $X$ is actually \textit{also} negative at solar system scale, therefore this procedure cannot be applied. The DW model therefore presents the same pathology as the RR nonlocal one, i.e. the remaining of a time dependence in the small scale limit of $G_{\rm eff}(k,z)$, Eq. (\ref{eq:limGeff}). 

In that case, the question reduces to asking if it is realistic to consider the limiting value $G_{\rm eff}(k,z)$, i.e. comprising FLRW background quantities, as valid in the solar system. Indeed, once the $k \longrightarrow \infty$ limit is taken on FLRW, one probes regions where the matter fluctuations becomes nonlinear and virialised, so linear perturbation theory on FLRW can in principle break down. Nevertheless, Ref.~\cite{Belgacem:2018wtb} argues that the linear perturbation theory on FLRW background, based on a metric expansion of the form of Eq.~(\ref{eq:interval}), is still valid at solar system scales. Hence, any of the models exposing the same pathology as the DW and the RR one are ruled out by LLR.

From our point of view, we believe that the conclusions of Ref.~\cite{Belgacem:2018wtb} are too strong in the view of the approximations made within their study. Indeed, here the problem consists in being able to understand how the FLRW background (i.e. averaged) quantities behave when evaluated from cosmological scales down to solar system ones, where the system ``decouples'' from the Hubble flow (such as within virialised objects). To do so, in contrast with respect to the method proposed  in Ref.~\cite{Belgacem:2018wtb}, we believe that a full non-linear time-- and scale--dependent solution around a non--linear structure would need to be studied. The screening properties in the DW and RR models should then be addressed in this framework. Lacking such a solution at the moment, we will not consider this issue any longer in this paper.

Nevertheless, if validated on conceptual ground, the conclusions of LLR constraints are severe for the DW model. Indeed, in Ref.~\cite{2014JCAP...09..031B}, the authors found that this quantity within the RR nonlocal gravity model is about $\dot{G}_{\rm eff}/G = 92 \, \times \, 10^{-13} \mathrm{yr^{-1}}$, putting the model under serious pressure. For the case of the DW model on its best fit to the CMB+SNIa+RSD data we find\footnote{This result quantifies the ``three orders of magnitude deviation'' estimated at the end of Sec.~4 of Ref.~\cite{Belgacem:2018wtb}.}, 
\begin{align}\label{eq:dotGeff}
\dot{G}_{\rm eff}/G = 3780 \, \times \, 10^{-13} \mathrm{yr^{-1}} \, ,
\end{align}
which strongly rules out the model in the view of the measurements of Ref.~\cite{Williams2004} at $\dot{G}_{\rm eff}/G = (4\pm9) \, \times \, 10^{-13} \mathrm{yr^{-1}}$, but again, provided the cosmological background solution remains valid on (very) small scales.

In any case, when seen from a cosmological perspective, the results of this section motivate the use of current, high precision complementary cosmological data from CMB observations, growth measurements from RSD and WL for constraining the DW nonlocal model. After presenting the relevant GR--deviations of the DW model in the tensor sector, such a study is addressed  in the rest of this article. 

\subsection{Indicators from GR--deviations: Tensor Sector}\label{sec:devGRtens}

The window of GW astronomy has recently been opened from the observations of several GW signals emitted from different coalescing binary systems in the sky. Most of the sources are thought to consist of black holes (see Refs.~\cite{Abbott:2016blz,Abbott:2016nmj,2041-8205-851-2-L35,Abbott:2017oio,Abbott:2017vtc}), but GWs from a binary neutron stars (BNS) merger has also been observed \cite{NSNSGRB2017}. The advantage of detecting BNS mergers is the potential of observing an associated electromagnetic counterpart in the IR, optical or UV wavelength. In that case, one can precisely determine the redshift of the source that fixes the redshift--mass degeneracy in the chirp mass, obtained from the time variation of the frequency of the signal. On the other hand, the detection of the strain amplitude of the GWs together with an estimation of the inclination angle of the binary system (that can for instance be obtained from polarisation--differentiated measurements) allows one to derive the luminosity distance, and therefore a Hubble diagram for GWs, making them ``standard sirens'' (see e.g. Refs.~\cite{Schutz:2001re,Holz:2005df} and Ref.~\cite{Maggiore:1900zz} for a standard textbook). One of such multi-messenger astronomical signals has been detected last year from the observations of the GWs from the binary neutron star merger GW170817 by the \textit{LIGO/Virgo} collaboration, as reported  in Ref.~\cite{NSNSGRB2017}, together with the (quasi--)simultaneous detection of the associated $\gamma$-ray burst GRB170817A, presented  in Refs.~\cite{Goldstein2017,Integral2017,Monitor:2017mdv}, and follow up studies of the electromagnetic counterpart in Ref.~\cite{GBM:2017lvd}. Accessing this measurement allowed to put strong constraints on the lower bound of the speed of GWs, which has been found to be equal to the speed of light within an error of $\mathcal{O}(10^{-15})$. This constraint implied dramatic restrictions for modified theories, such as those belonging to the (beyond) Horndeski class, for which several operators into the action are now forbidden (see e.g. Refs.~\cite{Lombriser:2015sxa, Lombriser:2016yzn,Baker:2017hug,Creminelli:2017sry,Ezquiaga:2017ekz,Sakstein:2017xjx}, Ref.~\cite{Ezquiaga:2018btd} for a review and ~\cite{2018JCAP...06..029A} for a possible way out), but also in nonlocal modified gravity, as shown in Ref.~\cite{Belgacem2017a}. 

Another source of deviations to GR in the tensor sector is the way gravitational waves are damped under the cosmic Hubble flow. Indeed, the usual propagation equation for tensor modes in GR is given by, 
 \begin{align}
& {h}_{ij}'' + 2 \hc {h}_{ij}' - \Delta h_{ij}= 16 \pi G \, a^2 \, \pi_{ij} \, .\label{eq:GWs}
\end{align}
where $\pi_{ij}$ is the traceless-tranverse (helicity--$2$) anisotropic stress matter tensor, non-vanishing when relativistic particles are present in the course of the evolution \cite{Durrer1997,Weinberg2003}. In going to Fourier space, writing $h_{ij}$ and $\pi_{ij}$ on a basis of helicity--2 polarisation tensors, e.g. $h_{ij} \equiv h_{A} Q_{ij}^A$, where $A= +, \times$ (see e.g. Ref.~\cite{Maggiore:1900zz}), redefining the fields as $h_{A}(\tau, k) \equiv \chi_A(\tau, k) / a(\tau)$ and discarding the source, one gets, 
\begin{align}
\chi_A'' + \bigg( k^2 - \frac{a''}{a} \bigg) \chi_A  = 0 \, . \label{eq:GWc}
\end{align}
At small scales, $k^2 \gg a''/a$ and the first term in the parenthesis dominates. As such, these equations show that the GWs propagate with dispersion relation $k^2 = \omega^2$ in GR, i.e. at the speed of light. For binary inspirals on a cosmological FLRW background, the amplitude of the GWs is related to the inverse of the luminosity distance to the source,
\begin{align}
h_A(\tau, k) \sim \frac{1}{D_L\big(z(\tau)\big)} \, , \qquad \qquad D_L(z) \equiv (1+z) \int^z_0 ~ \frac{\dd z'}{H(z')} \, .
\end{align}
Up to systematic factors, the precise knowledge of the amplitude of the GWs and redshift of the source therefore allows one to reconstruct a Hubble diagram from GW sources, making them ``standard sirens'' \cite{Schutz:2001re,Holz:2005df}. As a GW interferometry is independent from any experiments observing electromagnetic signals, such as the CMB, closeby or distant SNIa or galaxy cluster observations, their constraints are complementary. Regarding the expansion history, cosmological constraints from current observations of GWs still remain quite loose, as only one event with the electromagnetic counterpart has been recorded, \textit{viz}. GW170817/GRB170817A \cite{NSNSGRB2017}. However, GW observations prove to be very useful in the close future, especially within the light of third (or ``2.5'') generation GW interometers such as the \textit{LISA} \cite{LISA} and \textit{DECIGO} \cite{DECIGO} space telescopes or the ground--based \textit{Einstein Telescope} \cite{ET} and \textit{Cosmic Explorer} \cite{CE}. Examples are provided by the study in Ref.~\cite{Belgacem:2018lbp}, where the constraints on $H_0$ from GW170817/GRB170817A and 3G--detector forecast have been studied (see also the forthcoming, more realistic, Ref.~\cite{Belgacem2019:inprep}).

If GR is modified, the coefficient of the friction term $h_A'$ in Eq.~\eqref{eq:GWs}, can possibly be altered in the generic fashion (here we follow the discussion of Ref.~\cite{Belgacem:2018lbp}, but see also e.g. Refs.~\cite{Amendola:2014wma,Nishizawa:2017nef,Arai:2017hxj,Amendola:2017ovw,Ezquiaga:2018btd} for related discussions), 
 \begin{align}
& {h}_{ij}'' + 2 \hc \big( 1 - \Xi(\tau) \big)  {h}_{ij}' + k^2 h_{ij}= 0 \, ,\label{eq:GWmod}
\end{align}
which implies that the transformation to get rid of the Hubble friction term for obtaining Eq.~\eqref{eq:GWc}, now reads $h_A(\tau,k) \equiv \chi_A(\tau,k) / \tilde{a}(\tau)$, with, 
\begin{align}
\frac{\tilde{a}'}{\tilde{a}} \equiv \hc \big( 1 - \Xi(\tau) \big) \, .\label{eq:atilde}
\end{align}
This shows that, in such a class of models, GWs propagate at the speed of light and the above condition in turn implies that the strain of the GWs is now proportional to, 
\begin{align}
h_A \sim \frac{\tilde{a}(z)}{a(z)} \frac{1}{D_L(z)} \, , 
\end{align}
from which one defines the luminosity distance to GWs sources, 
\begin{align}
D^{\rm gw}_L(z) \equiv \frac{a(z)}{\tilde{a}(z)} D_L^{\rm em}(z) \, .
\end{align} 
From Eq.~\eqref{eq:atilde}, one can solve for $\tilde{a}(z)$ in terms of $\Xi(z)$ to obtain \cite{Belgacem2017a,Belgacem:2018lbp}, 
\begin{align}
D^{\rm gw}_L(z) \equiv  D_L^{\rm em}(z) \times \exp \bigg( - \int^z_0 \frac{\dd z'}{1+z'} \,  \Xi(z') \bigg) \, .\label{eq:DLGW}
\end{align} 
In modified gravity theories, the Hubble diagram built up from standard candles (i.e. electromagnetic sources) can therefore be different from the one constructed from standard sirens. Measuring the deviations between the Hubble diagram from electromagnetic and GW sources can therefore provide a clear signature for modification to GR. Such a fact has been studied in Ref.~\cite{Belgacem:2018lbp} which, rewriting Eq.~\eqref{eq:DLGW} by using a function parametrised by the pair $\big( \Xi_0, n \big)$ such as, 
\begin{align}
\frac{D_L^{\rm gw}(z)}{D_L^{\rm em}(z)} = \Xi_0  + a(z)^n \big(1- \Xi_0 \big) \, , \label{eq:ratioDL}
\end{align} 
found that the parameter $\Xi_0$ can be constrained four times more accurately than the equation of state today $w_0$ from the well--known Chevallier-Polarski-Linder parametrisation \cite{Chevallier2000,Linder2002}), given current CMB+SNIa+BAO constraints joined with forecast constraints from third generation GW experiments. The perspectives to probe modifications to gravity from future generation of GW intereferometers are therefore more optimistic than originally expected. Similar cosmological constraints including refined data analyses for the assumed coincident electromagnetic+GWs signals detection rate will be presented in Ref.~\cite{Belgacem2019:inprep}.   

\begin{figure}[h!]
\begin{center}
\vspace{-0.5cm}
\includegraphics[width=0.625\columnwidth]{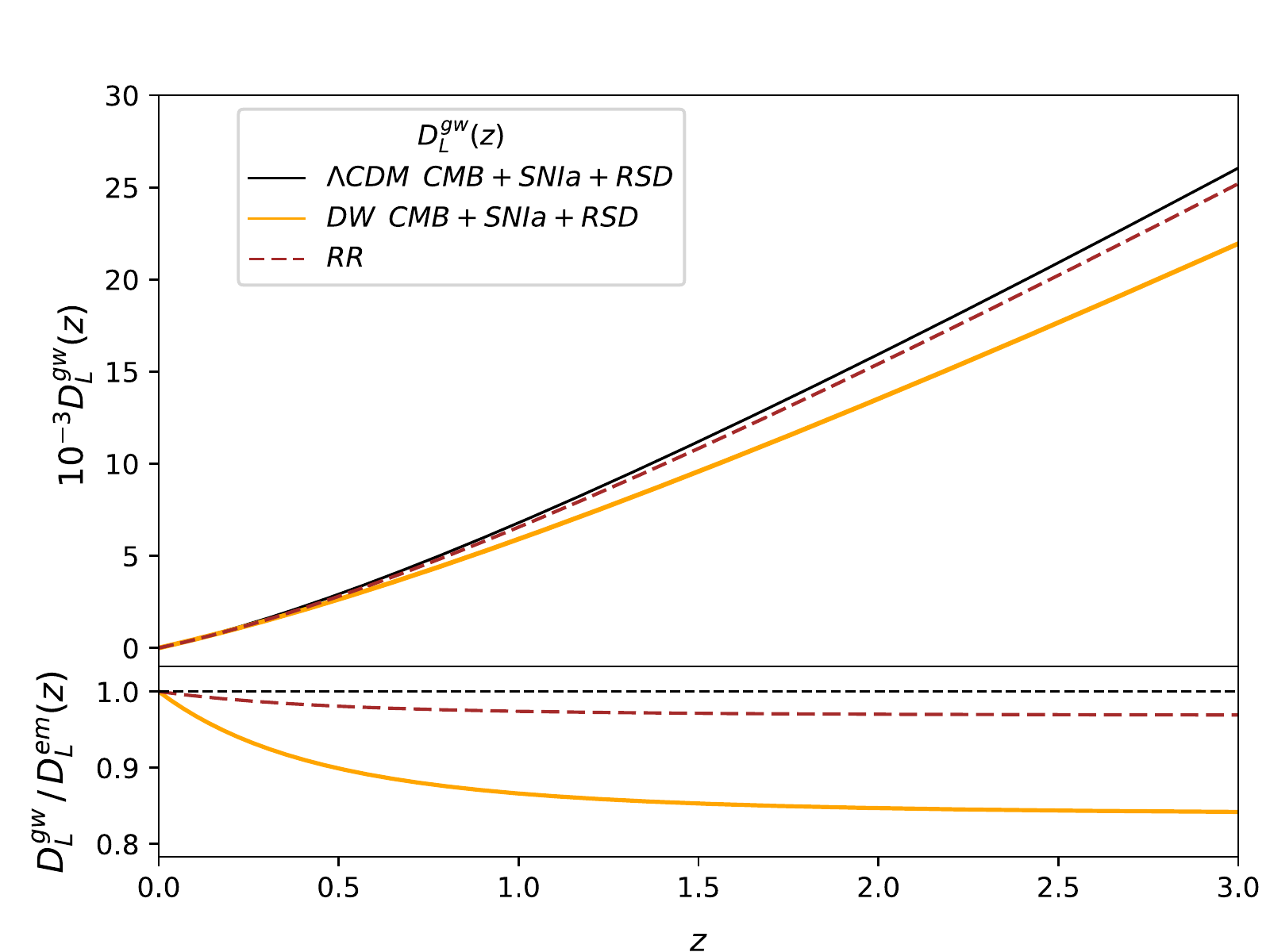}
\vspace{-0.15cm}
\caption{\label{fig:DLgw} \textit{Upper panel: }gravitational waves luminosity distance as a function of redshift in $\Lambda$CDM (black solid), DW (orange solid) on their best fit to CMB+SNIa+RSD and for the RR model (brown dashed). \textit{Lower panel:} the ratio between luminosity distance from standard sirens to the one from standard candles.}
\vspace{-0.5cm}
\end{center}
\end{figure}

In Sec.~\ref{sec:eqlin}, we have seen that the DW model features a similar modification in the Hubble friction term than in Eq.~\eqref{eq:GWmod}. Such a modification involves the quasi--static limit of the effective Newton's constant (see Eq.~\eqref{eq:GWprop}) and this structure is found in several theories of modified gravity \cite{Saltas:2014dha,Belgacem:2017ihm,Linder:2018jil}. Likewise, in the DW model, the modified luminosity distance for GWs, as compared to the one provided by electromagnetic sources, is given by, 
\begin{align}
D_L^{\rm gw}(z) \equiv D_L^{\rm em}(z) \times \sqrt{\frac{G_{\rm eff, gw}(z)}{G_{\rm eff, gw}(z=0)}} \, , \label{eq:ratioDD}
\end{align}
where $G_{\rm eff, gw}(z)/G \equiv \left. G_{\rm eff}(z, k)/G \right|_{|k|\gg1}$ and is seen from the orange solid curve in the left panel of Fig.~\ref{fig:Geffeta}. The luminosity distance for GWs is shown in the upper panel of Fig.~\ref{fig:DLgw}, while the lower panel shows its ratio to the electromagnetic luminosity distance, for the $\lcdm$ (black solid) and DW (orange solid) model on their respective bestfit to the CMB+SNIa+BAO data that we describe in Sec.~\ref{sec:data}. Following the parametrisation of Ref.~\cite{Belgacem:2017ihm} in Eq.~\ref{eq:ratioDL}, and fixing the cosmological parameters to the best fit to CMB+SNIa+RSD data, we find that the DW model provides the parameter values $\Xi_0 = 0.843$ and $n = 11/4 = 2.75$. These values can be compared to the preferred ones in the RR nonlocal model being $\Xi_0^{\rm RR}= 0.97$, $n^{\rm RR}=5/2$ \cite{Belgacem:2018lbp}, and whose luminosity distance to standard sirens is shown by the brown--dashed curves  in Fig.~\ref{fig:DLgw}\footnote{The cosmological parameter values used for the RR model are the best fitting values inferred from the constraints given CMB+SNIa+BAO data performed in Ref.~\cite{Dirian:2016puz}.}. 
In Ref.~\cite{Belgacem:2018lbp}, the authors worked out an estimation of the number of standard sirens needed to discriminate between standard $\lcdm$ and the RR nonlocal model, given current high precision CMB+SNIa+BAO data joined to the forecast constraints of third generation GW interferometers. Within the (although greatly simplified, see e.g. Ref.~\cite{Belgacem2019:inprep}) framework of their study, the authors find that $\sim 50$, $\sim 200$ and $\sim 400$ standard sirens are needed to distinguish between $\lcdm$ and RR in the ``optimistic'', ``realistic'' or ``pessimistic'' scenario, respectively. As we can see from the values of $\big( \Xi_0, n \big)$ for DW as compared to the ones of RR, the deviations from GR within the former are much more prominent than those in the RR model. This means that less standard sirens will be needed to distinguish between the $\lcdm$ model and DW. As GW interferometry (cosmological) pipelines are still under development (see e.g. Ref.~\cite{Belgacem2019:inprep}), a quantitative estimation of that number within our context is behind the scope of the present paper. Similarly, constraints on the cosmological parameter space in the DW, given the forecast sensitivity of future GWs experiments is left for future work.

In what follows, we carry out observational constraints and model selection given current complementary cosmological data.

\section{Observational Constraints \& Model Selection}\label{sec:MCMC}

We place observational constraints on the cosmological models made out of the DW nonlocal gravity and GR plus a cosmological constant $\Lambda$, given high precision complementary data from CMB, SNIa and RSD observations. In the previous section, we have displayed a set of relevant indicators to conveniently quantify the extent to which the DW model deviates from GR. On these grounds, we motivate the cosmological probes we use in the constraints and introduce the %considered 
associated datasets. Then, given different joined combinations of these datasets, we analyze the impact induced by the GR--deviations of the DW model on the preferred cosmological parameter subspace and compare it with the one preferred by $\lcdm$. Finally, we also apply (approximate Bayesian) model selection for comparing the $\lcdm$ and the DW models at each step. We then draw our statistical conclusion and discuss future perspectives.

\subsection{Datasets and Priors}\label{sec:data}

Here, we discuss the type of cosmological probes we choose to efficiently constrain the DW model and point out the associated observational datasets. In Sec.~\ref{sec:pheno}, we have shown several relevant indicators of deviations to GR in the scalar and tensor sectors of the theory. In the scalar sector, we have seen that the set $\{G_{\rm eff}, \mu, \Sigma, \eta \}$ in Eqs.~\eqref{eq:eta}--\eqref{eq:sig} predicted by both models, offered a convenient way to appreciate the deviations of the DW model to GR. For the same fiducial cosmology, the  DW model has a background cosmology similar to the one in $\lcdm$ and their deviations therefore only lie in their linear (and nonlinear) perturbations. We have seen that the DW model features a deficit of growth of linear structures -- about $-10 \%$ in the total matter power spectrum, at the scales of interest -- but a higher lensing power as compared to $\lcdm$. Both effects begin at late time, when the dark energy density starts to dominate the energy density of the Universe (at $z \lesssim 1$). 
\begin{table}[h!]
\centering
\resizebox{10.cm}{!}{
\begin{tabular}{|c|c|c|c||c|c|c|c|} 
 \hline 
z & $f \sigma_8$  & $\pm 1 \sigma_{f \sigma_8}$ & Survey & z &  $f \sigma_8$  & $\pm 1 \sigma_{f \sigma_8}$ & Survey\\  \hline 
0.001 &    0.505       &      0.085  &    2MTF ~\cite{Howlett:2017asq}							& 0.52  &    0.488       &      0.065  &    BOSS DR12~\cite{Wang:2017wia} \\
0.02  &    0.428       &      0.0465 &    6dFGS+SNIa~\cite{Huterer:2016uyq}			& 0.56  &    0.482       &      0.067  &    BOSS DR12~\cite{Wang:2017wia} \\
0.02  &    0.314       &      0.048  &    2MRS~\cite{Davis:2010sw,Hudson:2012gt} 							& 0.57  &    0.417       &      0.045  &    SDSS DR10 and DR11~\cite{Sanchez:2013tga} \\
0.02	 & 0.398          &   0.065     & SNIa+IRAS~\cite{Hudson:2012gt,Turnbull:2011ty}						& 0.57  			& 0.426  &  0.029  &    BOSS CMASS~\cite{Gil-Marin:2016wya}  \\
0.067 &    0.423       &      0.055  &    6dFGS~\cite{Beutler:2012px}							&  0.59  			&    0.481       &      0.066  &    BOSS DR12 ~\cite{Wang:2017wia}\\
0.1   &    0.370       &      0.130  &    SDSS-veloc~\cite{Feix:2015dla}					&  0.59  &    0.488       &      0.060  &    SDSS-CMASS~\cite{Chuang:2013wga} \\
0.1   &    0.48        &      0.16   &    SDSS DR13 ~\cite{Feix:2016qhh}				& 0.60  &    0.390       &      0.063  &    WiggleZ~\cite{Blake:2012pj} \\
0.1   &    0.376       &      0.038  &    SDSS DR7 	~\cite{Shi:2017qpr}					& 0.60  &    0.433       &      0.067  &    SDSS-BOSS~\cite{Tojeiro:2012rp} \\
0.15  &    0.490       &      0.145  &    SDSS-MGS ~\cite{Howlett:2014opa}					&  0.60  &    0.550       &      0.120  &    VIPERS PDR-2~\cite{Pezzotta:2016gbo} \\
0.17  &    0.510       &      0.060  &    2dFGRS ~\cite{Song:2008qt}						&  0.64  &    0.486       &      0.070  &    BOSS DR12~\cite{Wang:2017wia} \\
0.18  &    0.360       &      0.090  &    GAMA~\cite{Blake:2013nif}						& 0.727 &    0.296       &      0.0765 &    VIPERS~\cite{Hawken:2016qcy} \\
0.25  &    0.3512      &      0.0583 &    SDSS-LRG-200~\cite{Samushia:2011cs} 			& 0.73  &    0.437       &      0.072  &    WiggleZ~\cite{Blake:2012pj}  \\
0.3   &    0.407       &      0.055  &    SDSS-BOSS ~\cite{Tojeiro:2012rp}					& 0.76  &    0.440       &      0.040  &    VIPERS v7~\cite{Wilson:2016ggz}  \\
0.31  &    0.469       &      0.098  &    BOSS DR12~\cite{Wang:2017wia}  				& 0.77  &    0.490       &      0.18   &    VVDS~\cite{Song:2008qt} \\
0.32  &    0.427       &      0.056  &    BOSS-LOWZ~\cite{Sanchez:2013tga}  				&  0.80  &    0.470       &      0.08   &    VIPERS~\cite{delaTorre:2013rpa} \\
0.32  &    0.48        &      0.10   &    SDSS DR10 and DR11~\cite{Sanchez:2013tga}  &  0.85  &    0.45        &      0.11   &    VIPERS PDR-2~\cite{Mohammad:2017lzz}  \\
0.35  &    0.429       &      0.089  &    SDSS-DR7-LRG~\cite{Chuang:2012qt} 			& 0.86  &    0.48        &      0.10   &    VIPERS~\cite{delaTorre:2016rxm} \\
0.36  &    0.474       &      0.097  &    BOSS DR12~\cite{Wang:2017wia} 					&  0.86  &    0.400       &      0.110  &    VIPERS PDR-2~\cite{Pezzotta:2016gbo}  \\
0.37  &    0.4602      &      0.0378 &    SDSS-LRG-200~\cite{Samushia:2011cs} 			&  0.978 &    0.379       &      0.176  &    SDSS-IV~\cite{Zhao:2018jxv} \\
0.38  &    0.440       &      0.060  &    GAMA~\cite{Anderson:2013zyy} 							&   1.05  &    0.280       &      0.080  &    VIPERS v7~\cite{Wilson:2016ggz} \\
0.40  &    0.419       &      0.041  &    SDSS-BOSS~\cite{Tojeiro:2012rp} 					& 1.23  &    0.385       &      0.099  &    SDSS-IV~\cite{Zhao:2018jxv} \\
0.40  &    0.473       &      0.086  &    BOSS DR12~\cite{Wang:2017wia} 					& 1.40  &    0.482       &      0.116  &    FastSound~\cite{Okumura:2015lvp} \\
0.44  &    0.413       &      0.080  &    WiggleZ~\cite{Blake:2012pj} 							&  1.52  &    0.420       &      0.076  &    SDSS-IV~\cite{Gil-Marin:2018cgo} \\
0.44  &    0.481       &      0.076  &    BOSS DR12~\cite{Wang:2017wia} 					& 1.52  &    0.396       &      0.079  &    SDSS-IV~\cite{Hou:2018yny}  \\
0.48  &    0.482       &      0.067  &    BOSS DR12~\cite{Wang:2017wia} 					& 1.526 &    0.342       &      0.070  &    SDSS-IV~\cite{Zhao:2018jxv}  \\
0.5   &    0.427       &      0.043  &    SDSS-BOSS~\cite{Tojeiro:2012rp} 					& 1.944 &    0.364       &      0.106  &    SDSS-IV~\cite{Zhao:2018jxv}\\ 
\hline
\end{tabular}
}
\caption{\label{tab:fs8} Redshifts, means, standard deviations and names with corresponding references of the various RSD measurements we consider in that work.}
\end{table}
\\
We therefore need cosmological surveys that efficiently probe the growth history of the Universe. To date, robust data are provided by growth features such as the linear growth rate $f \sigma_8$, accurately measured by RSD observations, as well as the WL (of e.g CMB or galaxies), that usually measure the combination $\sim \sigma_8 (\Omega_M)^\alpha$, where $\alpha$ is a number depending on the survey considered. To that end, we constrain the linear growth of structures with several measurements of $f \sigma_8$, from various surveys that resolve RSD (see Table~\ref{tab:fs8} for details), while the lensing power is constrained with the (lensed) CMB temperature and polarisation (cross) spectra, as well as the tri--spectrum extracted CMB WL measurements of \textit{Planck} 2015 \cite{Planck2015CP}.

For linear cosmological perturbations, the CMB power spectra calibrate the amplitude of the matter power spectrum through the determination of $\sigma_8$ within a given model, while RSD measurements form an additional and complementary probe. Additionally, at the background level, the CMB constrains the acoustic distance--scale ratio $\theta_* \equiv r_s(z_*)/D_A(z_*)$ with great accuracy and in turn fixes the value of the sound horizon to recombination, given additional data are provided to break the geometrical degeneracy between $\Omega_M$ and $\Omega_\Lambda$ in $D_A(z_*)$ (see e.g. Ref.~\cite{Efstathiou1998}). For instance, one could make use of the galaxy clustering BAO for building up a distance ladder for constraining further the expansion history at late time. However, geometrical distortions between longitudinal and transversal BAOs (Alcock-Palschinsky effect), induced by e.g. modifications of gravity, are degenerated with RSD. Therefore, for all the RSD measurements that we consider in that work (see Table~\ref{tab:fs8}), we would need to access and implement the full covariance matrix that includes BAO shape measurements for controlling that degeneracy. Since such a requirement is hard for us to realise in practice, we use RSD measurements where the BAO shape information has been marginalised out. Although this is a $\lcdm$--dependent procedure, it is not worrisome in our case since the DW model describes the same FLRW background as $\lcdm$. Moreover, we see from Fig.~\ref{fig:mPk} that, around the scale of interest for the RSD ($\sim 100 {\, \rm Mpc}$), no scale dependence is introduced in the growth by the DW modification to GR, as the matter power spectrum solely undergoes a constant shift. This is also true for the redshift range of interest for RSD measurements, as we have explicitly checked. We are therefore safe to use current RSD data to constrain the DW model. 

Lacking the BAO shape information associated with these RSD data, we adopt the simplest (but not less worth) strategy to fix the angular distance to the last scattering surface by using distance scale data from the distant SNIa of the \textit{JLA} compilation described in Ref.~\cite{JLA2014}. This partially breaks the geometrical degeneracy from CMB independent data and makes the inferred constraints especially robust. 

Direct $H_0$ measurements which prefer higher values as compared to \textit{Planck} CMB (as the ones mentioned in Sec.~\ref{sec:intro}) will exert the same tension within the DW model as compared to within $\lcdm$, as both have the same background expansion history. 

Finally, we notice that the combination of the amplitude of fluctuation $\sigma_8$ and density fraction $\Omega_M$ of matter, can also be efficiently probed by observations of the cosmic shear from galaxy WL or from X--ray or SZ selected cluster counts. However, as mentioned  in Sec.~\ref{sec:intro}, these data are still subject to current debates about systematic issues that makes the interpretation of the resulting constraints delicate. We therefore make the conservative choice not to include such probes into our analysis, but nevertheless discuss their potential impact at the end of our study.

\noindent We now present the datasets we use in the observational constraint study that follows.  

\paragraph*{\hspace{-0.35cm}\textbf{CMB.}}

For the CMB, we use the likelihoods of \textit{Planck} 2015 \cite{Planck_2015_1} from measurements of the (cross-) power spectra of the CMB.
In particular, we take the lowTEB data for low multipoles ($\ell \leq 29$) and the high-$\ell$ Plik  TT,TE,EE (cross-half-mission) ones for the high multipoles ($\ell > 29$) of the power spectra \cite{Planck2015DE, Planck_2015_wiki}. Moreover, in order to further constrain the excess of lensing in the DW model, we also include the power spectrum of the lensing potential extracted from the CMB trispectrum (where only the conservative multipole range $\ell =40-400$ is used). Such a consideration allows to break degeneracies in the primary CMB anisotropies and to place further constraints on the growth history at late times (see e.g. \cite{Planck_2015_Lkl, Planck_2015_lens}). In what follow, this set of CMB data will be quoted as \textit{Planck} and the same dataset excluding the reconstructed lensing maps as \textit{Planck wo lensing}.   

\paragraph*{\hspace{-0.5cm} \textbf{Distant SNIa.}}

For the distant SNIa, we consider the data of the \textit{SDSS-II/SNLS3} Joint Light-curve Analysis (JLA) of Ref.~\cite{JLA2014}, and make use of the complete (non-compressed) corresponding likelihoods.  

\paragraph*{\hspace{-0.5cm} \textbf{RSD.}}

Finally, the RSD measurements that we consider are listed  in Table~\ref{tab:fs8}. To use this data, we implement the computation of the growth rate, 
\begin{align}
f \sigma_8(z) = \frac{\dd \log \delta_M}{\dd \log a} \sigma_8(z) \, , 
\end{align}
where the perturbation quantities are evaluated at the wave number of interest for RSD surveys, i.e. $k \simeq 10^{-2} {\, \rm Mpc^{-1}}$. We then construct the RSD likelihood in considering the data points provided in Table~\ref{tab:partable}. In our constraints, we compress these data in forming weighted averages of points close in redshift, i.e. in bins of $\Delta z \simeq 0.1$. 

\paragraph*{\hspace{-0.35cm}\textbf{Priors}.} Regarding the cosmological setup we use, the primordial fluctuations are considered to be adiabatic and Gaussian, with a slightly red tilted power spectrum, together with which, the reionisation history and matter content in the Universe are taken according to the \textit{Planck} baseline (see e.g. Ref~\cite{Planck_2013_CP}). The prescription for $f$ in the DW model we consider, i.e. reproducing the $\lcdm$ expansion history  (see Sec.~\ref{sec:funcf} for more details), precisely requires the input of the $\lcdm$ Hubble expansion rate $H_{\lcdm}(z)$, so the DW model has the same number of parameters as $\lcdm$.  Within the baseline, the models can be parametrized by the six--dimensional vector, 
\begin{align}
\theta_{\rm base} = \big( H_0,\, \omega_b,\, \omega_c,\, \ln ( 10^{10} A_s) ,\, n_s,\, \tau_{\rm re} \big) \, , \label{eq:base_param}
\end{align}
where $H_0$ is the Hubble parameter today,  $\omega_b \equiv \Omega_b h^2$ and $\omega_c \equiv \Omega_c h^2$ are the physical baryon and cold dark matter density fractions today, respectively,  $A_s$ is the amplitude and $n_s$ the spectral tilt of the power of primordial scalar perturbations, and $\tau_{\rm re}$ is the reionization optical depth. We choose improper flat priors on all these parameters, except for $\tau_{\rm re}$ which is bounded from below at $0.01$, in accordance with Gunn-Peterson trough observations (see e.g. Ref.~\cite{Becker:2001ee}). The baseline also assumes the presence of two massless neutrinos together with a third one which is massive. The massive neutrino mass therefore plays the role of the absolute neutrino mass $M_\nu$ and is fixed to the lowest value allowed by terrestrial experiments within the baseline, that is, $M_\nu=0.06 \,{\rm eV}$ \cite{Planck_2013_CP}. However, first, considering a varying $M_\nu$ is perfectly legitimate given the present bounds from neutrino oscillation experiments (see e.g. Sec.~F of Ref.~\cite{Dirian2017} and Refs. therein) and second, the absolute neutrino mass is known to be degenerated with effects of modified gravity at background, linear and nonlinear perturbation level (see for example Refs.~\cite{2014MNRAS.440...75B, Bellomo:2016xhl, Dirian2017}), and fixing the prior $M_\nu=0.06 \,{\rm eV}$, could therefore result in biased inferences. This is for instance the case in particular classes of modified gravity theories, as for the RR nonlocal model considered in Refs.~\cite{Dirian:2016puz,Dirian2017}, where the authors show that a dark energy of the phantom nature, i.e. with equation of state $w_{\textsf{de}}(z) < -1$, can typically give raise to a tension between \textit{Planck} CMB, \textit{JLA} distant SNIa and RSD data, in the framework of the baseline Eq.~(\ref{eq:base_param}). Nevertheless, the authors then also show that this discrepancy can be healed by allowing the absolute neutrino mass $M_\nu$ to vary. For the DW model, the background cosmology is the same as in $\lcdm$ and the growth of linear structures is lower as compared to the one in $\lcdm$, thus fixing $M_\nu$ to its lowest allowed value is, a priori, sufficient in the present framework for consistently confronting the DW model against $\lcdm$.
\begin{table}[h!]
\centering
\resizebox{15cm}{!}{
\begin{tabular}{|l||c|c||c|c||c|c|} 
 \hline 
\multicolumn{1}{|l||}{ } & \multicolumn{2}{|c||}{\textit{Planck}} & \multicolumn{2}{|c|}{\textit{Planck}+RSD} & \multicolumn{2}{|c|}{\textit{Planck}+RSD+\textit{JLA}} \\ \hline
Param & $\Lambda$CDM  & DW & $\Lambda$CDM & DW  & $\Lambda$CDM & DW\\ \hline 
$100~\omega_{b }$  &
$2.225_{-0.016}^{+0.016}$ & 
$2.232_{-0.018}^{+0.018}$ &
$2.230_{-0.018}^{+0.018}$ &
$2.220_{-0.018}^{+0.017}$ &
$2.231_{-0.019}^{+0.018}$ &
$2.222_{-0.017}^{+0.017}$ \\

$\omega_c$  & 
$0.1194_{-0.0015}^{+0.0014}$ &  
$0.1186_{-0.0017}^{+0.0015}$ & 
$0.1187_{-0.0016}^{+0.0016}$ & 
$0.1202_{-0.0015}^{+0.0016}$ & 
$0.1186_{-0.0016}^{+0.0015}$ &  
$0.1200_{-0.0014}^{+0.0015}$ \\ 

$H_0$  &
$67.5_{-0.66}^{+0.65}$&  
$67.87_{-0.72}^{+0.73}$ & 
$67.79_{-0.74}^{+0.75}$ & 
$67.19_{-0.71}^{+0.66}$ &
$67.82_{-0.73}^{+0.71}$ &
$67.28_{-0.69}^{+0.61}$\\ 

$\ln (10^{10}A_{s })$ & 
$3.064_{-0.025}^{+0.025}$ &
$3.010_{-0.029}^{+0.025}$ & 
$3.051_{-0.028}^{+0.029}$ & 
$3.009_{-0.029}^{+0.023}$ &
$3.052_{-0.028}^{+0.029}$ & 
$3.011_{-0.028}^{+0.025}$ \\ 

$n_{s}$  & 
$0.9647_{-0.0049}^{+0.0048}$ &  
$0.9661_{-0.0054}^{+0.0050}$  &  
$0.9657_{-0.0055}^{+0.0053}$ & 
$0.9628_{-0.0053}^{+0.0049}$ & 
$0.9657_{-0.0055}^{+0.0054}$ &  
$0.9633_{-0.0047}^{+0.0049}$ \\ 

$\tau_{\rm re}$  & 
$0.0653_{-0.014}^{+0.014}$  & 
$0.03957_{-0.016}^{+0.013}$ & 
$0.06008_{-0.015}^{+0.016}$ & 
$0.03748_{-0.015}^{+0.013}$ &  
$0.06041_{-0.016}^{+0.015}$ & 
$0.03862_{-0.015}^{+0.013}$ \\ 
\vspace{-0.2cm}
\,&\,&\,&\,&\,&\,&\\
$\sigma_8$  & 
$0.8171_{-0.0089}^{+0.0089}$ & 
$0.7530_{-0.0096}^{+0.008}$ &  
$0.8098_{-0.0094}^{+0.0098}$ &
$0.7583_{-0.0088}^{+0.0076}$ & 
$0.8098_{-0.001}^{+0.0099}$ & 
$0.7584_{-0.0085}^{+0.0082}$ \\

$z_{\rm re}$ & 
$8.752_{-1.2}^{+1.4}$ & 
$5.999_{-1.6}^{+1.7}$ &  
$8.194_{-1.4}^{+1.7}$ & 
$5.8_{-1.6}^{+1.6}$ & 
$8.222_{-1.4}^{+1.6}$ & 
$5.929_{-1.5}^{+1.7}$\\ 
   
\hline
$\chi^2_{\rm RSD}$ &$-$ &$-$&$22.6$ &$31.3$  &$22.2$ &$29.1$\\
$\Delta \chi^2_{\rm RSD}$ &$-$ & $-$&$0$ &$8.7$  &$0$ &$6.9$\\
\hline 
$\chi^2_{JLA}$ &$-$ &$-$&$-$ &$-$  &$341.5$ &$342.0$\\
$\Delta \chi^2_{JLA}$ &$-$ & $-$&$-$ &$-$  &$0$ &$0.5$\\
\hline 
$\chi^2$ &$12943.3$ &$12943.4$&$12967.1$ &$12972.4$  &$13650.3$ &$13655.7$\\
$\Delta \chi^2$ &$0$ & $0.1$&$0$ &$5.3$  &$0$ &$5.4$\\
$\Delta B$ &$0$ & $1.6$&$0$ &$5.2$  &$0$ &$5$\\
$\alpha$ & $\times$ & $1.01$ & $\times$ & $1.82$  & $\times$ &$1.77$\\
\hline
\end{tabular}
}
\caption{\label{tab:partable} Means and standard deviations of the inferred cosmological parameters given the associated dataset and (effective) $\chi^2$ goodness-of-fit. The $\Delta\chi^2$ values are taken with respect to the lowest value within each dataset and $\chi^2 \equiv -2 \, {\rm ln} \, \mathcal{L}$, where $\mathcal{L}$ is the likelihood function. The quantity $\Delta B$ and $\alpha$ are defined in Eq.~\eqref{eq:deltaB} and Eq.~\eqref{eq:alpha}, respectively.}
\vspace{-0.cm }
\end{table}

\subsection{Results}\label{sec:res} 

In this section, we perform observational constraints on the DW and $\lcdm$ models given three joined combinations of the high precision complementary datasets presented %here 
above.
\\
\begin{figure}[t]\label{fig:ClTT}
\vspace{-0.cm}
\centering
~~~~~~~~\includegraphics[width=\columnwidth]{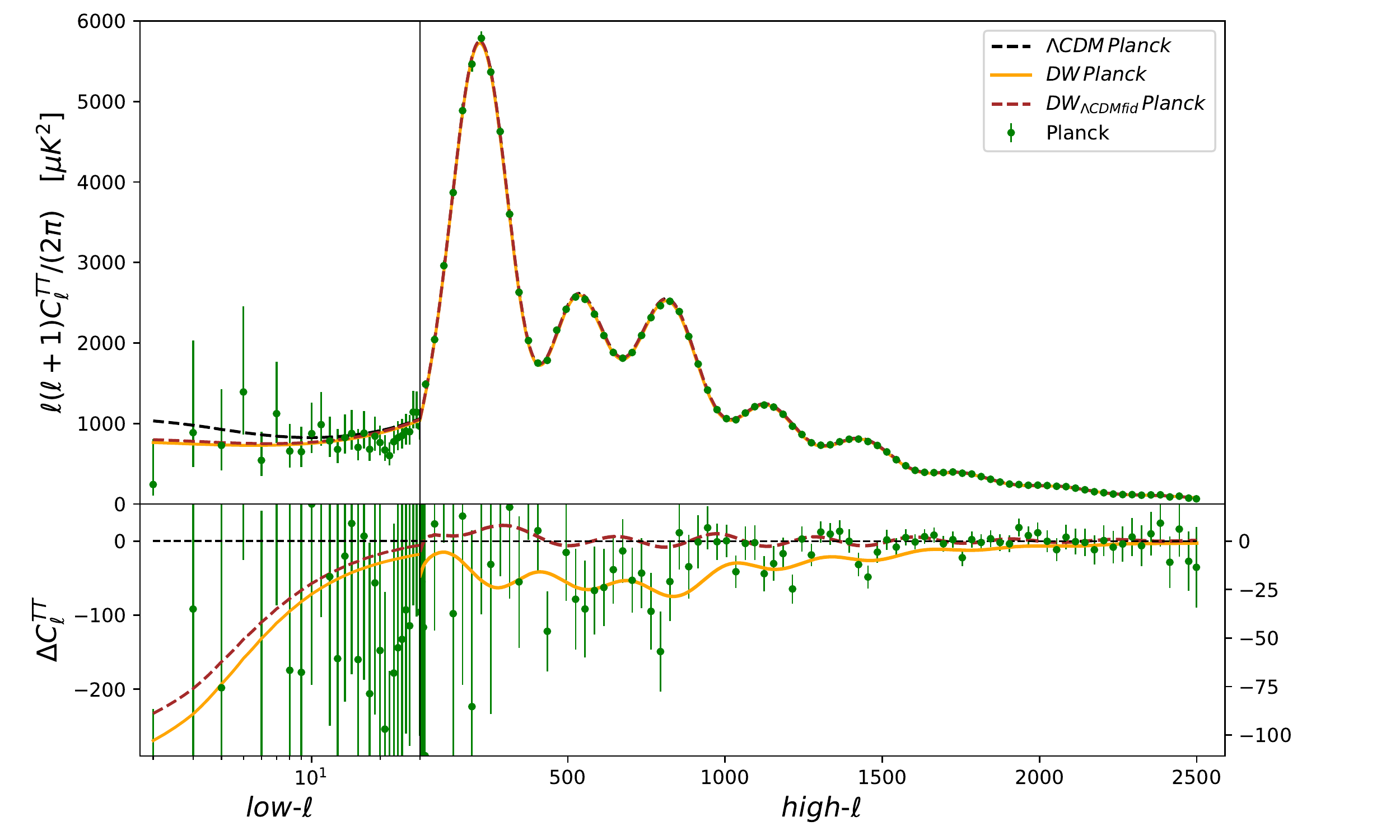} 
\vspace{-0.5cm}
\caption{\label{fig:ClTT} \textit{Upper panel:} temperature power spectrum for the $\Lambda$CDM (black dashed), the DW (orange solid) on their best fit to \textit{Planck} CMB and for the $\dwlcdm$ model (brown dashed).
\textit{Lower panel:} residuals for $\Lambda$CDM and the difference between the prediction in DW (orange solid) and $\dwlcdm$ (brown dashed) as compared to the one of $\Lambda$CDM. Green data points are from the \textit{Planck} 2015 release \cite{Planck2015CP}. Error bars correspond to $\pm 1 \sigma$ uncertainty.
}
\vspace{-0.cm}
\end{figure}
To begin with, we start by evaluating the posterior distribution within the six--dimensional parameter space Eq.~\eqref{eq:base_param}, provided the priors and the \textit{Planck} CMB dataset described in Sec.~\ref{sec:data}.
The second and third columns of Table~\ref{tab:partable}, show the inferred cosmological parameter means and standard deviations for the $\lcdm$ and the DW model, respectively. As we can see, the background--related parameters $\big\{ H_0,\, \omega_b,\, \omega_c \big\}$, do not significantly change from $\lcdm$ to DW ($\lesssim 0.5 \sigma$ shift). Of course, this results from the fact that the DW model under consideration is designed so as to reproduce the $\lcdm$ cosmological background history. The mild shifts in the background--parameters are mostly due to correlations with other parameters or statistical fluctuations. 
Nevertheless,  among the quantities mostly related to the linear perturbations $\big\{ \ln ( 10^{10} A_s) ,\, n_s,\, \tau_{\rm re} \big\}$, $A_s$ and $\tau_{\rm re}$ are significantly different from the preferred ones in $\lcdm$. 
\begin{figure}[h!]
\centering
\includegraphics[width=0.495\columnwidth]{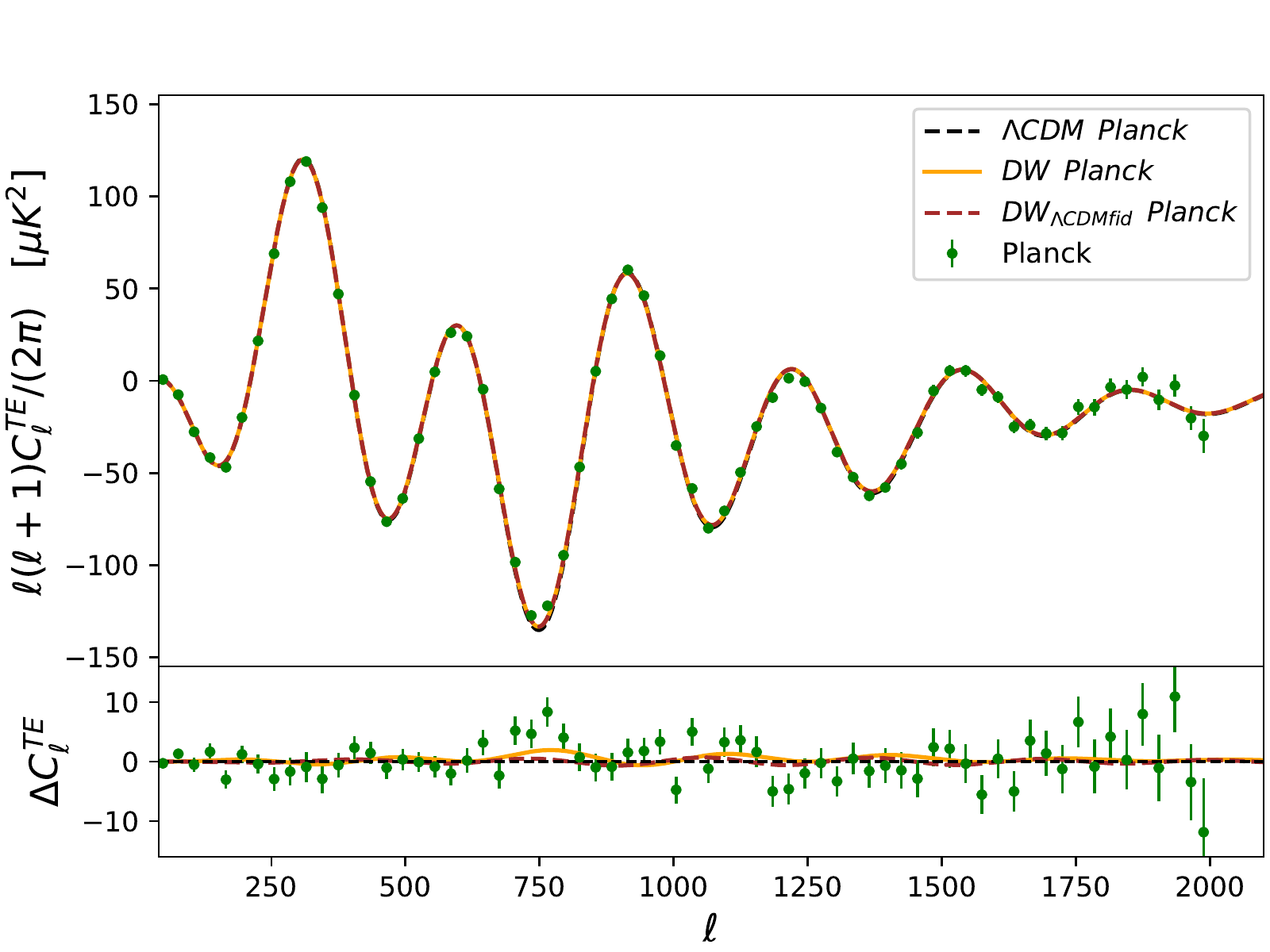} 
\includegraphics[width=0.495\columnwidth]{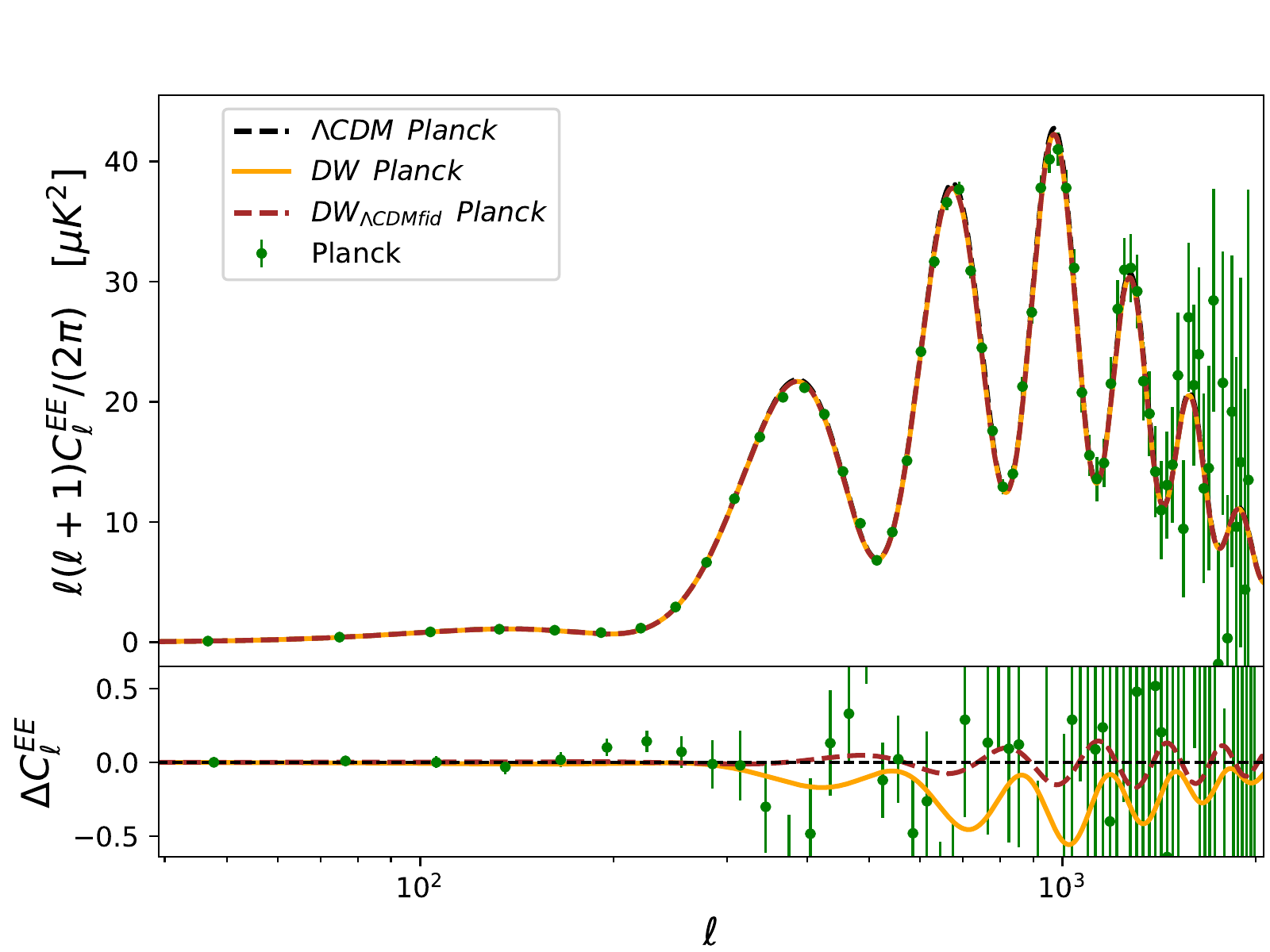} 
\vspace{-0.25cm}
\caption{\label{fig:ClTEEE} \textit{Upper left panel:} TE cross--correlation power spectrum for the $\Lambda$CDM (black dashed) and the DW (orange solid) on their best fit to \textit{Planck} CMB, and for the $\dwlcdm$ model (brown dashed).
\textit{Lower left panel:} residuals for $\Lambda$CDM and corresponding difference in the TE--power spectra. \textit{Upper right panel:} E--mode CMB polarization power spectrum for the $\Lambda$CDM (black dashed) and the DW (orange solid) on their best fit to \textit{Planck} CMB, and for the $\dwlcdm$ model (brown dashed). \textit{Lower right panel:} residuals for $\Lambda$CDM and the corresponding difference in the EE--power spectra. Data points are from \textit{Planck} 2015 \cite{Planck2015CP} (green bars). Error bars correspond to $\pm 1 \sigma$ uncertainty.
}
\vspace{-0.cm}
\end{figure}

We show the angular power spectrum of the CMB temperature anisotropies predicted from the $\lcdm$, DW and $\dwlcdm$\footnote{Recall this is the DW model on $\lcdm$--best fitting  cosmological parameters to the data under consideration.} models and compare them with \textit{Planck} 2015 measurements in Fig.~\ref{fig:ClTT}. The left panel of Fig.~\ref{fig:ClTEEE} shows the same for TE cross correlation power spectrum while the right panel of the same figure shows the EE angular power spectrum. We can see that there is a strong reduction of the power at low--$\ell$ in the TT power for $\dwlcdm$. This region being dominated by the integrated Sachs--Wolfe (ISW) effect, this results from a lower gravitational potential $\Psi$ at late-time and can be seen in Fig.~\ref{fig:mu}. Moreover, on higher $\ell$'s in all the spectra, one can see a mismatch in the prediction of the CMB peaks and troughs amplitude in the $\dwlcdm$ model. We see that peaks are lower and troughs less deep, i.e. the lensing smoothing of the CMB temperature and polarisation (cross) power spectra is increased within the DW model. Indeed, this agrees with the results of Sec.~\ref{sec:pheno} [see e.g the indicator $\Sigma(z,k)$ in Fig.~\ref{fig:sig}], where we saw that the DW model describes a higher lensing power as compared to $\lcdm$, for fixed cosmological parameters. This means that for accessing a given amount of lensing smoothing for the CMB TT, TE or EE power spectra, or of the reconstructed lensing potential, the amplitude of the fluctuations ($\sim A_s \sim \sigma_8$) therefore needs to be smaller in DW as compared to $\lcdm$. This explains why the amplitude of primordial fluctuations $A_s$ significantly shifts, or equivalently $\sigma_8$ that undergoes a $\sim 7 \sigma$ lower shift, and in turn also why the TT, TE and EE power spectra are lowered once the DW model is fit to \textit{Planck} CMB (see Figs.~\ref{fig:ClTT} and~\ref{fig:ClTEEE}).

\begin{figure}[h!]
\vspace{0.15cm}
\centering
\includegraphics[width=0.625\columnwidth]{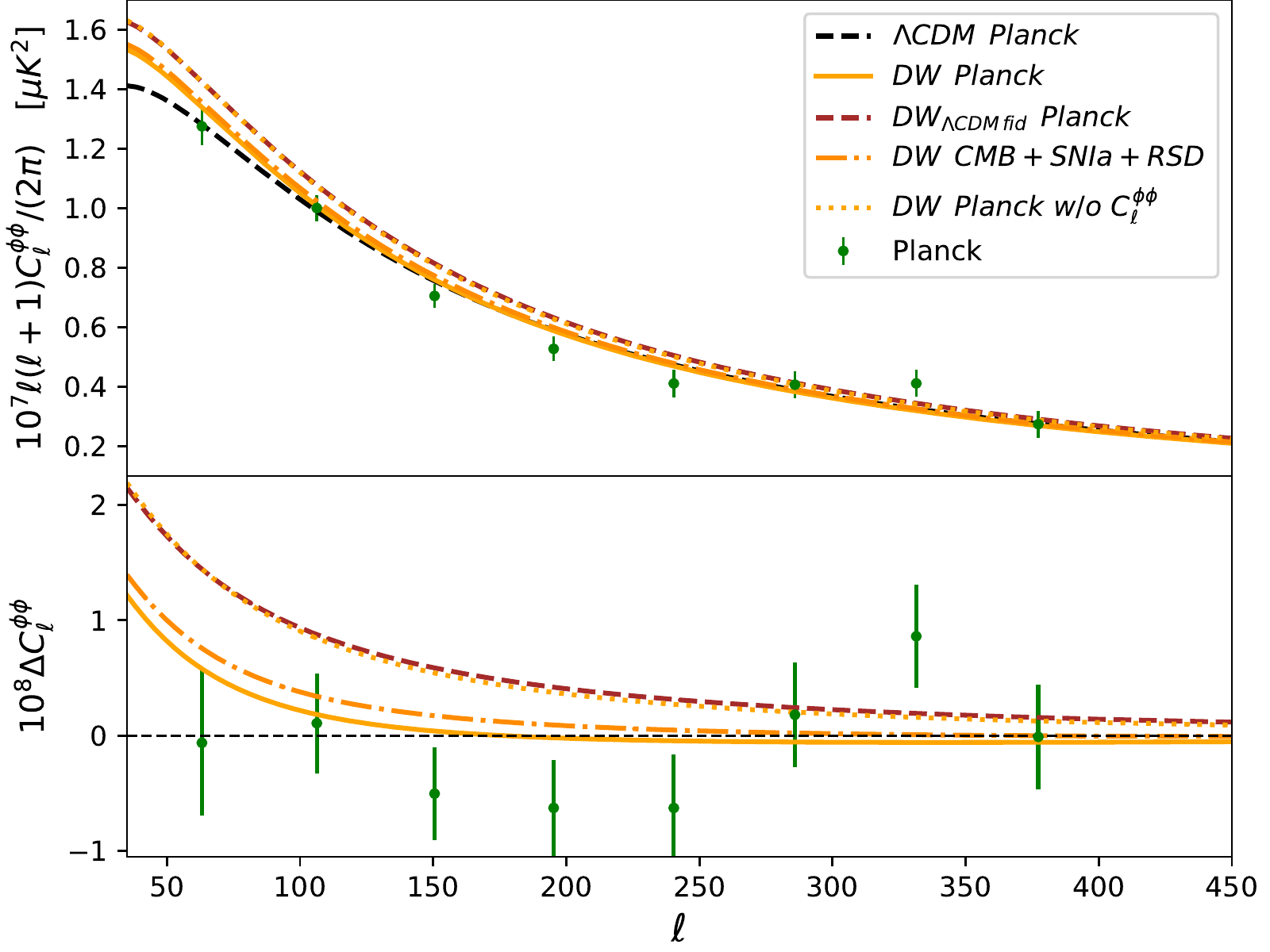}
\vspace{-.25cm}
\caption{\label{fig:ClFF} \textit{Upper panel:} lensing potential power spectrum in $\Lambda$CDM (black dashed), DW (orange solid) and $\dwlcdm$ (brown dashed) given \textit{Planck} 2018 data. We also display the predictions of DW given joined CMB+SNIa+RSD data (orange dot--dashed) and \textit{Planck w/o lensing} (orange dotted).
\textit{Lower panel:} residual data points for $\Lambda$CDM and the difference between the prediction in DW (orange solid) and in $\dwlcdm$ (brown dashed) with respect to the one in $\Lambda$CDM. Data points are from \textit{Planck} 2015 \cite{Planck2015CP} (green bars). Error bars correspond to $\pm 1 \sigma$ uncertainty.
}
\end{figure}

\begin{figure}[h!]
\centering
\includegraphics[width=0.625\columnwidth]{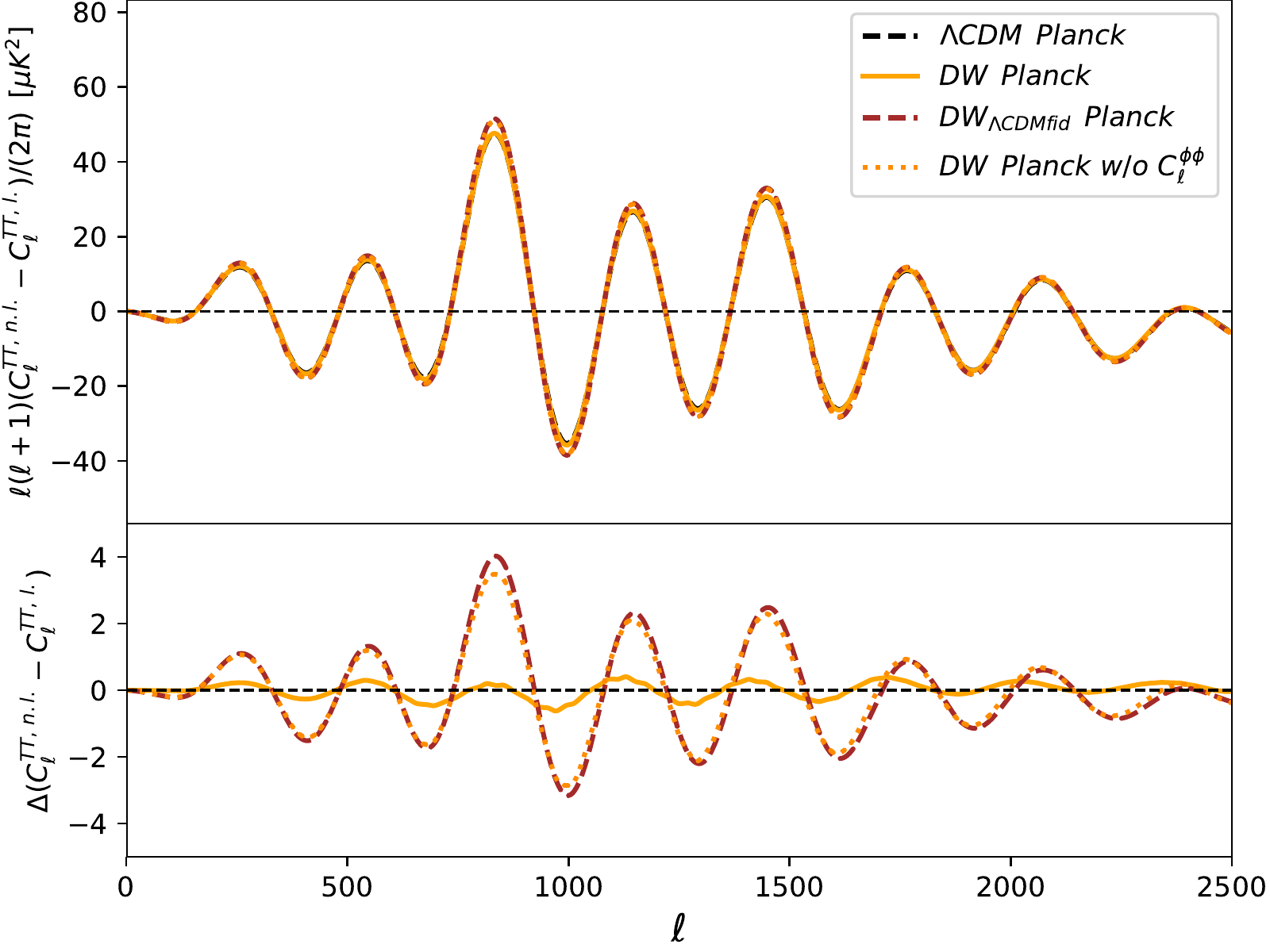} 
\vspace{-.25cm}
\caption{\label{fig:ClTTdiff}
\textit{Upper panel:} differences between the unlensed (ul.) and lensed (l.) temperature angular power spectra for the same models as in Fig.~\ref{fig:ClFF}.
\textit{Lower panel:} differences of $C^{TT, \, ul.}_{\ell} - C^{TT, \, l.}_{\ell}$ between the prediction of the DW model (orange solid) and of $\dwlcdm$ (brown dashed) with respect to the one in $\Lambda$CDM.
}
\end{figure}

The excess of lensing power within the DW model is better illustrated in Fig.~\ref{fig:ClFF} which displays the CMB lensing potential power spectrum $\ell ( \ell + 1)C_\ell^{\phi \phi}$ and Fig.~\ref{fig:ClTTdiff} that shows the difference between the unlensed (ul.) and the lensed (l.) CMB TT angular power spectrum, as predicted from the $\lcdm$, DW and $\dwlcdm$ models. From the latter, we can indeed see that the lensing smoothing to the CMB TT power spectrum in the DW model is $\sim +10\%$, when the DW model parameters are fixed to the $\lcdm$ \textit{Planck} best fit. 

\begin{figure}[h!]
\centering
\includegraphics[width=0.5\columnwidth]{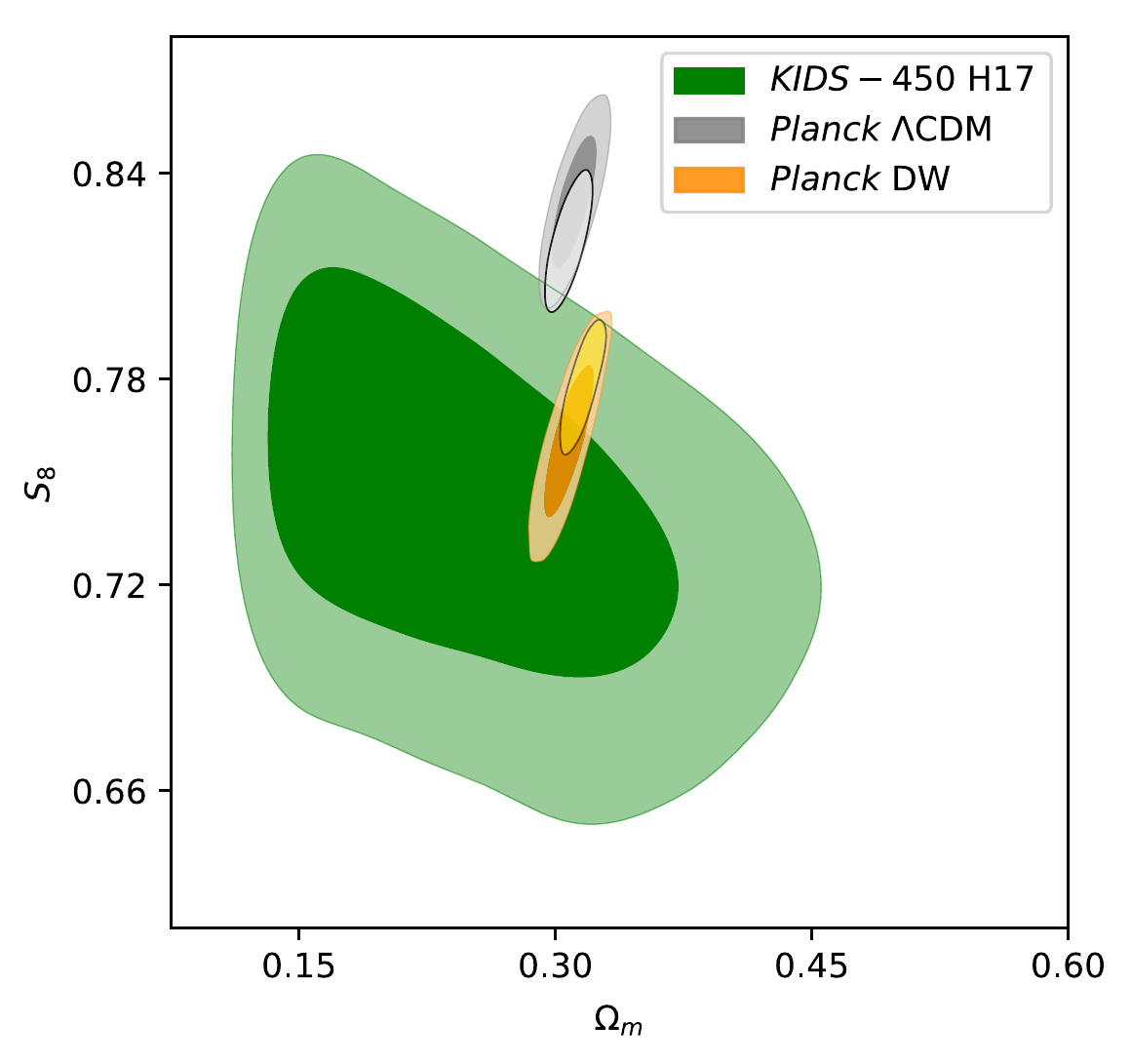} 
\vspace{-0.5cm}
\caption{\label{fig:triWL}Two dimensional marginalised posterior distribution in the $\Omega_M$--$S_8$ plane for DW (orange contour) and $\lcdm$ (grey contour) provided \textit{Planck} CMB data and individual constraints from the \textit{KIDS-450} survey of Ref.~\cite{Hildebrandt:2016iqg} (green contour). The shaded regions correspond to $1\sigma$ and $2\sigma$ confidence level. The corresponding black encircled lighter regions correspond to the $1\sigma$ contour given the CMB+SNIa+RSD data we describe  in Sec.~\ref{sec:data}.}
\vspace{-0.35cm}
\end{figure}

On Figs.~\ref{fig:ClFF} and~\ref{fig:ClTTdiff}, we also show the prediction of the DW model given \textit{Planck} CMB data when excluding the reconstructed $C_\ell^{\phi \phi}$. We can see that the lensing smoothing of the TT power spectrum is also $\sim +10 \%$ stronger in that case. This is caused by the fact that the CMB power spectra generically prefer stronger lensing smoothing (i.e. overestimes $\sigma_8$) than the weak lensing potential extracted from the temperature anisotropies four--point function (see Fig.~\ref{fig:ClFF}, and e.g. Refs.~\cite{Planck2015CP,Aghanim:2018eyx} for more details). Such a fact implies that the addition of CMB WL $C_\ell^{\phi \phi}$ data to the fit to \textit{Planck} CMB TT+TE+EE power spectra decreases the amplitude of matter fluctuations.
Moreover, Fig.~\ref{fig:triWL} clearly displays the distinction between the predictions of the DW and $\lcdm$ models in the $\Omega_M$--$S_8 [\equiv \sigma_8 (\Omega_M/0.3)^{0.5}]$ plane. Another way to understand why the amplitude of matter fluctuation is smaller in DW is to refer to the left panel of Fig.~\ref{fig:Geffeta}, where we can see that the effective Newton constant $G_{\rm eff}(z)$ is significantly increased at small redshifts  in the DW model as compared to $\lcdm$, as well as to the RR nonlocal model. 

In turn, as the damping tail of the CMB accurately measures the combination $\sim A_s e^{-2 \tau_{\rm re}}$, together with a decrease in the amplitude $A_s$, the optical depth to reionisation $\tau_{\rm re}$ also decreases (i.e. the smaller the fluctuations, the later the reionisation). The spectral tilt $n_s$ is however almost unaffected.

From the two last lines of the second and third columns of Table~\ref{tab:partable}, we can read the $\chi^2$ goodness-of-fit obtained from both distributions\footnote{The sampling method for obtaining such values and associated parameters bestfits is detailed  in Ref.~\cite{Dirian:2016puz}.}. We can see that, despite having very different behaviour in their linear scalar perturbations, and therefore different preferred cosmological parameter values, both models are statistically indistinguishable given the \textit{Planck} CMB data, as they have almost equal $\chi^2$ goodness-of-fit values.
\begin{figure}[h!]
\vspace{-0.5cm}
\centering
\includegraphics[width=0.75\columnwidth]{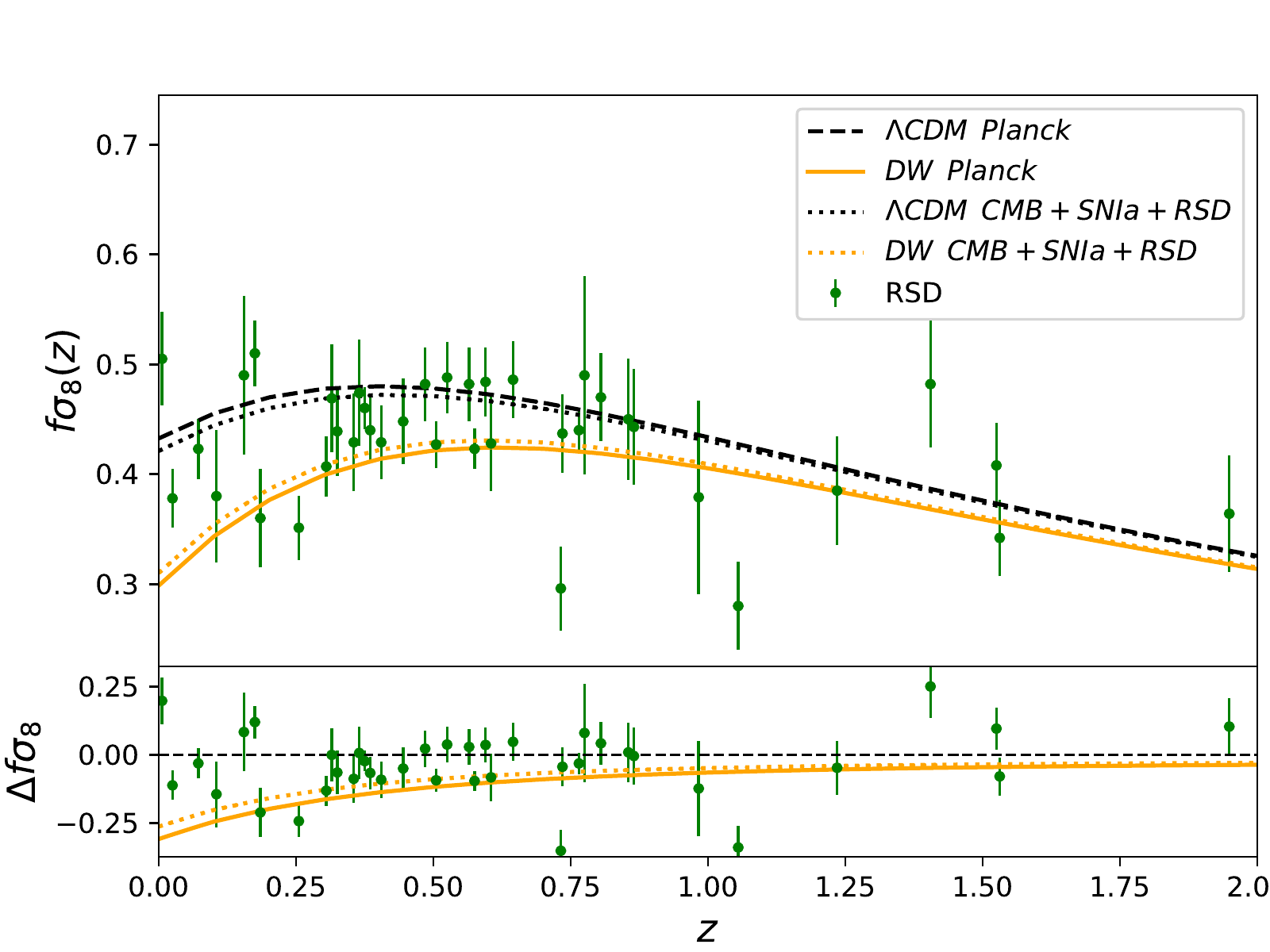}
\vspace{-0.15cm}
\caption{\label{fig:fs8}\textit{Upper panel: }growth rate computed for the $\Lambda$CDM (black dashed) and the DW (orange solid) on their \textit{Planck} CMB bestfit, and for the $\dwlcdm$ model (brown dashed), together with the data points detailed in Table.~\ref{tab:fs8}. \textit{Lower panel:} corresponding residuals and relative differences.}
\vspace{-0.25cm}
\end{figure}

As discussed in Sec.~\ref{sec:data}, we use the RSD measurements reported in Table~\ref{tab:fs8} and displayed in Fig.~\ref{fig:fs8}, to further constrain the CMB--calibrated growth history described by the DW model, and compare its performance to the one of standard $\lcdm$ given \textit{Planck} CMB+RSD data. The results are shown  in the fourth and fifth columns of Table~\ref{tab:partable}. Concerning parameter shifts, we can see that the addition of RSD data to \textit{Planck} CMB has the opposite effects on the models. Indeed, while $\sigma_8$ is pushed towards slightly lower values in $\lcdm$ ($\sim 0.5 \sigma$), it is preferred higher in the DW model. This fact can also be been in the $\Omega_M$--$S_8$ plane, as illustrated by the two extra lighter $1\sigma$ contours in Fig.~\ref{fig:triWL}, that are produced using CMB+SNIa+RSD data\footnote{As will be discussed below, the addition of distant SNIa data from Ref.~\cite{JLA2014} does not affect our argument on growth and lensing properties of the DW model.}.
This fact can also be understood in referring to Fig.~\ref{fig:fs8}, where we can see that the \textit{Planck}--CMB calibrated growth rate described by the DW model underestimates most of the $f \sigma_8$ data values from RSD. The data therefore favour a higher amplitude $\sigma_8$, so as to compensate for the too large deficit in the growth. On the contrary, $\lcdm$ slightly overestimates the data and the RSD constraints therefore favour lower amplitudes to matter fluctuations than \textit{Planck} CMB data alone. As a consequence, because the amplitude of fluctuations is anti-correlated with the total matter density fraction $\Omega_M$, the latter tends to (although quite slightly) increase  in the DW model, while it tends to decrease within $\lcdm$. Furthermore, as the CMB shape information accurately constrains the combination $\omega_M \equiv \Omega_M h^2$, the Hubble constant $H_0$ therefore increases accordingly in $\lcdm$, and decreases within the DW model, a (not significant) tendency that is however not preferred by the direct measurements of $H_0$ evoked in Sec.~\ref{sec:intro}.

From the two last lines of the fourth and fifth columns of Table~\ref{tab:partable}, we see that the addition of RSD measurements to \textit{Planck} CMB data creates a tension within the DW model as compared to $\lcdm$ (with $\Delta \chi^2_{\rm RSD}=8.7$ and $\Delta \chi^2=5.3$).
We observe that the appearance of this tension is mostly caused by the preference of the DW model for stronger CMB lensing features but milder growth. Indeed, when joining the RSD measurements to \textit{Planck} CMB within DW, the trend for lower $\sigma_8$ induced by boosted CMB lensing features of the DW model, in addition to a milder lensing power favoured by the reconstructed \textit{Planck} CMB WL data, then competes with the preference for higher growth of the RSD data, i.e. for a larger $\sigma_8$. Reciprocally, such a fact unavoidably comes together with an increased CMB lensing power. In the DW model, this tends to push the predicted CMB lensing potential $C_\ell^{\phi \phi}$ away from \textit{Planck} CMB $1\sigma$ errorbars at low--$\ell$, as can be seen from the left panel of Fig.~\ref{fig:ClFF}. This is precisely where the CMB--RSD tension is at play.

The standard $\lcdm$ model turns out to be favoured over DW with a Bayes Information Criterion (BIC) (see e.g. Ref.~\cite{Trotta2008} for a comprehensive discussion) of $\Delta \chi^2 = 5.3$, given \textit{Planck} CMB+RSD data. This value becomes $\Delta \chi^2 =5.4$, when also including the SNIa for refining further the constraints. According to the Jeffrey scale reported  in Ref.~\cite{Trotta2008}, such BICs can be interpreted as ``weak'' evidences for $\lcdm$ against the DW model given the data. The conclusion is that the joined dataset described in Sec.~\ref{sec:data}, does not possess enough constraining power for significantly distinguishing between standard $\lcdm$ and DW.  

A better approximation to the Bayesian factor $B_{12}$, which encapsulates a simplified version of the  Occam's razor, has been proposed in Ref.~\cite{deAlmeida:2018kwq} and reads,
\begin{equation}
 2 \ln B_{12}\approx \Delta B=\Delta \chi^2 - \ln \frac {P_1F_2}{P_2 F_1}
\end{equation}
where $P_{1,2}, \,  F_{1,2}$ are the determinants of the prior and of the likelihood parameter inverse covariance matrices (i.e. of the Fisher matrices), respectively, of the two models $1,2$ to compare. In our case, the two models have the same number of parameters and the same priors, so $P_1=P_2$. Then, if model 1 is $\Lambda$CDM and model 2 is DW,
\begin{equation}
 \Delta B=\Delta \chi^2 - \ln \frac {F_2}{F_1} \, . \label{eq:deltaB}
\end{equation}
Since $B_{12}$ is to be interpreted as the odds of model 1 with respect to model 2, the expression
\begin{equation}
P_{12}=\frac{B_{12}}{1+B_{12}} \, ,
\end{equation}
is the probability that model 1 is the correct one, in a space of models represented only by 1 or 2. Then,
\begin{equation}
P_{12}=\frac{1}{1+e^{-\Delta B/2}} \, , 
\end{equation}
and Jeffrey's scale can be replaced by the usual $1,2,3\sigma$ probability levels of $P_{12}$. That is, 
model 1 is better than model 2 at a $\sigma$ level of, 
\begin{equation}
\alpha =-\sqrt{2}{\rm Erfc}^{-1} [\frac{1}{1+e^{-\Delta B/2}}+1] \, . \label{eq:alpha}
\end{equation}
In the framework of our study, the corresponding values of $\Delta B$ are found in the second last column of Table~\ref{tab:partable} and can be compared with the BIC $\Delta \chi^2$. We can see that the contribution of the supplemental term  in Eq.~\eqref{eq:deltaB}, are non--trivial but not significant as compared to the contributions of the BIC $\Delta \chi^2$, found in our case. 

The associated values of $\alpha$ are reported in the last line of Table~\ref{tab:partable}. Interpreted in such terms the standard $\lcdm$ model is favoured against the DW model as $\sim 1.82 \, \sigma$ given the  \textit{Planck} CMB+RSD data described in Sec.~\ref{sec:data}, while the discrepancy reduces to $\sim 1.77 \, \sigma$, when the \textit{JLA} SNIa lightcurve compilation is added to the latter. As already discussed, this discrepancy is not stringent enough for ruling out the DW model against $\lcdm$ on firm statistical grounds, and additional data are needed for potentially being able to tell the difference between both cosmologies.

\section{Future Perspectives \& Conclusions}\label{sec:conc}

We study the cosmological phenomenology of a particular class of nonlocal modification to gravity provided by the DW nonlocal gravity model given in Eq.~\eqref{eq:DWaction}. Within this class, we consider a particular type of models where the free distortion function $f$ is fixed so as to reproduce a given $\lcdm$ expansion history within the DW model. Such a fact implies that the DW model only deviates from GR at the cosmological perturbation level that we have within the linear scalar and tensor sectors. 

In the former case, by using a set of relevant indicators, we have seen that the DW model generically describes a lower linear growth rate but a stronger lensing power, as compared to $\lcdm$ on the same cosmological parameter values.\\
Within the tensor sector, we saw (see Sec.~\ref{sec:devGRtens}) that the DW model modifies the way GWs are damped when propagating on a cosmological background, and this fact makes the Hubble diagram that can be constructed from standard sirens deviate from the one obtained from standard candles (i.e. electromagnetic sources). Such a deviation is known to be efficiently constrained by future third generation GW experiments and related forecast constraints are left for future work.

\smallskip

In Sec.~\ref{sec:MCMC}, we perform observational constraints and use model selection techniques to confront the DW model against standard $\lcdm$, given high precision cosmological data. Provided the studied GR--deviations induced by the DW model, we choose to constrain both cosmological models by using the (lensed) \textit{Planck} CMB temperature and polarisation (auto and cross) power spectra together with the tri--spectrum extracted CMB WL potential power spectrum. In addition, we join complementary RSD data to the CMB ones for exerting further constraints on the linear growth, as well as SNIa data for further increasing the constraining power. 

As a result, the DW model is statistically equivalent to $\lcdm$ in the light of the \textit{Planck} CMB dataset, but prefers significantly lower values for the amplitude of matter fluctuations $\sigma_8$, as well as for the optical depth to reionisation $\tau_{\rm re}$. We have argued that such a fact results from the higher lensing power in DW as compared to $\lcdm$, that pushes $\sigma_8$ to lower values. This trend is however not favoured by the linear growth rate data from RSD measurements we consider (see Table~\ref{tab:fs8}), as most of the $f \sigma_8$ data points lie above the CMB--calibrated prediction of the DW model (see Fig.~\ref{fig:fs8}). The RSD data therefore prefer higher values for $\sigma_8$ and this explains the mechanism driving the tension appearing within the DW model when joining \textit{Planck} CMB together with RSD data.

According to the Jeffrey scale, such a tension (with $\Delta \chi^2 \simeq 5.35$ or $\sim 1.77 \sigma$, see Sec.~\ref{sec:res}) is qualified as ``weak'' evidence against the DW model as compared with $\lcdm$. Thus, the cosmological data considered is not accurate enough to discriminate between the $\lcdm$ and DW models, and complementary information is needed for eventually being able to tell the difference. 

\smallskip

On the contrary, lower values of $\sigma_8$ are favoured by the galaxy WL data from the \textit{KiDS-450} survey. Indeed, in referring to the two--dimensional marginalised constraints in $\Omega_M$--$S_8$ plane Fig.~\ref{fig:triWL}, we see that the lower value of $S_8 \sim \sigma_8$ preferred by the DW model makes it more consistent  with the galaxy WL data from \textit{KiDS-450}, as compared with $\lcdm$. This implies that adding \textit{KiDS-450} WL data to our global fit would tend to favour the DW model over $\lcdm$, and therefore to pull the BIC towards negative values (i.e. where DW is favoured by the data) . However, as mentioned  in Sec.~\ref{sec:intro}, it is not yet clear that these data are free from uncontrolled systematics, so we decide to only briefly and qualitatively comment on their potential impact on our results, rather than including them within the full analysis. 

\smallskip

Moreover, in Sec.~\ref{sec:devGRtens}, we have seen that DW modifies the way GWs propagate as compared to $\lcdm$ and such a feature is known to be efficiently probed by future third generation GW interferometers (see e.g. Ref.~\cite{Belgacem:2018lbp}). Furthermore, the characteristic linear growth and lensing features of the DW model expose the need to be further constrained by current galaxy clustering and WL data, as for instance the ones from \textit{DES} \cite{Abbott:2017wau}, or future galaxy and WL surveys such as \textit{Euclid}, \textit{DESI}, \textit{LSST}, \textit{SKA} \cite{SKA1,SKA2} or \textit{Stage-4 CMB} experiments.
Future experiments such as galaxy and WL surveys, or GW interferometers therefore appear of prime interest for efficiently constraining the modifications to GR induced by the DW model, and for ultimately being able to distinguish it from standard $\lcdm$.  

\smallskip

Finally, we should not forget that this analysis is legitimate only in the case where one shows that the small-scale limit of the effective Newton constant $G_{\rm eff}(z)$ within the DW model, tends toward $G$ in bound objects (virialised systems). Indeed, when evaluating the effective Newton constant in DW in the subhorizon limit $\left. G_{\rm eff}(z,k) \right|_{|k| \gg 1}$, a FLRW background time--dependence persists within the latter and $G$ is not reached. However, FLRW background quantities only make sense on cosmological scales, and the nature of their behaviour when transposed to solar system scales is not clear at all. 

If one can show that $\left. G_{\rm eff}(z,k) \right|_{|k| \gg 1}=G$, the model possesses a screening mechanism that makes it agree with solar system constraints. However, if this is not the case, the model is severely ruled out by Lunar Laser Ranging constraints on the time variation of the Newton constant, as the prediction of the DW model given in Eq.~\eqref{eq:dotGeff} lies significantly off from the current errorbars.

\vspace{5mm}
\noindent
{\bf Acknowledgments.}
We are grateful to Michele Maggiore for stimulating discussions and thank the anonymous referee for useful suggestions.\\
The work of YD is supported by the Fonds National Suisse and by a Consolidator Grant of the European Research Council (ERC-2015-CoG grant 680886). The work of SP was supported by the European Research Council under the European Union's Seventh Framework Programme (FP7/2007-2013)/ERC Grant No. 617656, ``Theories and Models of the 
Dark Sector: Dark Matter, Dark Energy and Gravity''.\\
The numerical computations presented in this publication were carried out on the Baobab cluster of the University of Geneva.\\
The development of Planck has been supported by: ESA; CNES and CNRS/INSU-IN2P3-INP (France); ASI, CNR, and INAF (Italy); NASA and DoE (USA); STFC and UKSA (UK); CSIC, MICINN and JA (Spain); Tekes, AoF and CSC (Finland); DLR and MPG (Germany); CSA (Canada); DTU Space (Denmark); SER/SSO (Switzerland); RCN (Norway); SFI (Ireland); FCT/MCTES (Portugal).
A description of the Planck Collaboration and a list of its members, including the technical or scientific activities in which they have been involved, can be found at 
http://www.cosmos.esa.int/web/planck/planck-collaboration.

\bibliographystyle{utphys}
\bibliography{DWmodel.bib}

\end{document}